\documentclass[12pt]{iopart}

\usepackage{iopams}  
\usepackage{graphicx}  
\usepackage{epsf}
\usepackage{bm}
\begin{document}

\title[Regular second order perturbations]
{Regular second order perturbations of binary black holes:
The extreme mass ratio regime}

\author{Carlos O. Lousto and Hiroyuki Nakano}

\address{Center for Computational Relativity and Gravitation, 
School of Mathematical Sciences, 
Rochester Institute of Technology, Rochester, New York 14623, USA}
\ead{colsma@rit.edu, hxnsma@rit.edu}
\begin{abstract}
In order to derive the precise gravitational waveforms 
for extreme mass ratio inspirals (EMRI), 
we develop a formulation for the second order metric perturbations 
produced by a point particle moving in the Schwarzschild spacetime. 
The second order waveforms satisfy a wave equation with an 
effective source build up from products of the first order perturbations 
and its derivatives. We have explicitly 
regularized this source at the horizon and at spatial infinity. 
We show that the effective source does not contain squares of the Dirac's 
delta and that perturbations are regular at the particle location. 
We introduce an asymptotically flat gauge for the radiation fields and the
$\ell=0$ mode to compute explicitly the (leading) second order $\ell=2$
waveforms in the headon collision case. This case represents the first completion
of the radiation reaction program self-consistently.
\end{abstract}

\submitto{\CQG}
\maketitle

%%%%%%%%%%%%%%%%%%%%%%%%%%%%%%%%%%%%%%%%%%%%%%%%%%%%%%%%%%%%%%%%%%%%%%
\section{Introduction}
%%%%%%%%%%%%%%%%%%%%%%%%%%%%%%%%%%%%%%%%%%%%%%%%%%%%%%%%%%%%%%%%%%%%%%

In the past twenty years we have witnessed steady increase in the
interest in gravitational waves from astrophysical
sources. Specially driven by the design and construction of laser
interferometric detectors, both ground and space
based. Alongside with this experimental developments theoretical
progress has also been steady.  We are now in conditions to predict the
gravitational radiation from the astrophysical scenarios expected to
produce the strongest signals, i.e. the merging binary black holes.

The two main scenarios involving black holes are, first, galactic
binaries with black holes having comparable masses (a few solar
masses). They are, for instance, the product of a supernova
explosion plus a subsequent accretion. The second scenario involves a
supermassive black holes (with several million solar masses) residing
in the center of an active galaxy. They attract stars in the inner
nuclei towards unstable orbits with a subsequent plunge generating
observable gravitational radiation. This scenario clearly involves
extreme mass ratio collisions. Let us also mention that a less common
event, but most energetic, is the close collision of galaxies and
consequently of supermassive black holes in their respective cores.

From the theoretical point of view one advantage of dealing with
binary black holes is that one can treat the problem of generation of
radiation in terms of only its gravitational field, ignoring the
(small) effects of matter around the binary system. A second important
feature is that the equations of General Relativity scale with the
total mass of the system. In this way, one can choose the dimensions
of the systems a posteriori, i.e. after solving for the scale free
problem (See \cite{Lousto:2007ji} for the case of three black holes.). 
It is then convenient to characterize the binary black hole
systems in terms of the mass ratio of its components (besides the
individual spins, orbital parameters and spatial orientation with
respect to the observer).

In binary black hole systems most of the generation of gravitational
radiation take place during the final few orbits before the
merger. In the case of comparable masses, this stage involves highly
nonlinear interactions among black holes and can only be described by
directly solving numerically the full General Relativity field
equations. Until recently this represented an insurmountable task that
hold the field for nearly thirty years, but during the year 2005 two
successful approaches
\cite{Pretorius:2005gq,Campanelli:2005dd,Baker:2005vv} have lead to
stable codes that allow to simulate binary black holes in
supercomputers. This breakthrough in Numerical Relativity lead to
numerous studies during the last year, including the last few
quasi-circular orbits ~\cite{Campanelli:2006gf,Baker:2006yw}, the
effect of ellipticity \cite{Pretorius:2006tp}, and spin-orbit coupling
leading to a change in the merger time \cite{Campanelli:2006uy},
corotation \cite{Campanelli:2006fg} and spin-flip and precession
\cite{Campanelli:2006fy}. Finally, unequal mass black holes
have been studied in
\cite{Campanelli:2004zw,Herrmann:2006ks,Baker:2006vn,Gonzalez:2006md}
reaching a minimum mass ratio of nearly 1:4. But simulations with mass
ratios up to 1:10 are currently underway. Generic binary simulations,
i.e. unequal masses and spins, have first been reported in \cite{Campanelli:2007ew},
this simulation lead to the shocking discovery that merging spinning black
holes can acquire recoil velocities up tp 4000 km/s \cite{Campanelli:2007cga}.
Even multi-black hole spacetimes are now possible to evolve numerically 
\cite{Campanelli:2007ea,Lousto:2007rj}.

In the small mass ratio regime the smaller hole orbiting the larger is
considered as a perturbation. In this approach the smaller hole is
described by a Dirac's delta particle and the spacetime is no longer
empty but has a non-vanishing energy-momentum tensor at the location of
the particle. The simplicity of this approach is appealing, but the
problem notably complicates when self-force effects are taken into
account to compute the correction to the background geodesic motion. A
consistent approach to this problem has been first laid down 
ten years ago by Mino,
Sasaki and Tanaka \cite{Mino:1996nk} and later confirmed by Quinn and Wald
\cite{Quinn:1996am} by providing a regularization procedure. 
(See \cite{Poisson:2003nc,Poisson:2004gg} for a detailed review 
and \cite{Barack:2001gx} for a practical regularization method.)
Self-force corrections can be considered second order effects on
the mass ratio of the holes, the natural perturbation parameter
for the system. To consistently compute the gravitational waveform, 
radiation energy, angular and
linear momentum radiated to infinity (and onto the horizon) we need
to proceed to solve the second order perturbations problem for the
gravitational field. The formalism to study second (and higher) order
perturbations of rotating black holes with sources was extensively
discussed in \cite{Campanelli:1998jv} based on the Newman-Penrose approach
of curvature perturbations.

The first explicit computation of the gravitational self-force was
done for the headon collision of two black holes in
Ref. \cite{Lousto:1999za,Barack:2002ku}.  This allow us to complete here,
for the first time, the second order program applied to the headon
collision of black holes. 
Note that this step is also crucial in order
to close the gap between the full numerical simulation with comparable
masses and the perturbative approach for extreme mass ratios.

Here, the second order calculation is required 
to derive precise gravitational waveforms 
which are used as templates for gravitational wave data analysis. 
In general, this second order computation has to be done by numerical integration. 
It is hence important to derive a well-behaved second 
order effective source, and we will focus on this problem in this paper. 
The second order analysis was pioneered by Tomita~\cite{Tomita:1974,Tomita:1976}, 
and vacuum perturbations in the Schwarzschild background 
was studied by Gleiser {\it et al}.
~\cite{Gleiser:1995gx,Gleiser:1996yc,Gleiser:1998rw,Nicasio:2000ge}. 
There are also studies on second order quasi-normal modes~\cite{Ioka:2007ak,Nakano:2007cj}, 
and the second order analysis was also extended to 
cosmology~\cite{Mukhanov:1996ak,Matarrese:1997ay,Acquaviva:2002ud,Nakamura:2004rm,Tomita:2005et}.

This paper is organized as follows. 
We consider second order metric perturbations 
and the equations they satisfy, i.e., the perturbed Hilbert-Einstein equations
of General Relativity in \sref{sec:SOMP}. 
In \sref{sec:1st}, we discuss the first order metric perturbations 
in the case of a particle falling radially into a Schwarzschild black hole 
by using the Regge-Wheeler-Zerilli formalism~\cite{Regge:1957td,Zerilli:1971wd}. 
This formalism in the time domain is summarized in \ref{app:RWZ}. 
To calculate the second order source, 
the first order $\ell=0$ mode is given in an appropriate gauge in \ref{app:SOGT}. 
In \sref{sec:2nd}, we derive the regularized second order effective source 
where the singular behaviors are removed analytically. 
In Sec.~\ref{sec:dis}, we summarize this paper 
and discuss some remaining open problems. 
Details of the calculations are given in the appendices. 
Throughout this paper, we use units in which $c=G=1$. 

%%%%%%%%%%%%%%%%%%%%%%%%%%%%%%%%%%%%%%%%%%%%%%%%%%%%%%%%%%%%%%%%%%%%%%
\section{Second order metric perturbations}\label{sec:SOMP}
%%%%%%%%%%%%%%%%%%%%%%%%%%%%%%%%%%%%%%%%%%%%%%%%%%%%%%%%%%%%%%%%%%%%%%

We consider second order metric perturbations 
on black hole backgrounds, 
\begin{eqnarray}
\tilde g_{\mu\nu}=g_{\mu\nu}+h_{\mu\nu}^{(1)}+h_{\mu\nu}^{(2)} \,, 
\label{eq:g-exp}
\end{eqnarray}
with expansion parameter $\mu/M$ corresponding to the mass ratio of the holes. 
Here, $g_{\mu\nu}$ is the background metric, 
and superscripts $(i)$ ($i=1,\,2$) denote the perturbative order, 
i.e., $h_{\mu\nu}^{(1)}$ and $h_{\mu\nu}^{(2)}$ are called the 
first and second order metric perturbations, respectively. 
In the perturbative calculation, we raise and lower all tensor indices 
with the background metric. 
The Hilbert-Einstein tensor up to 
the second perturbative order is given by 
\begin{eqnarray}
G_{\mu\nu}[\tilde g_{\mu\nu}] &= 
G_{\mu\nu}^{(1)}[h^{(1)}]+G_{\mu\nu}^{(1)}[h^{(2)}]+G_{\mu\nu}^{(2)}[h^{(1)},h^{(1)}] 
\,, 
\end{eqnarray}
where we have omitted the spacetime indices $\mu$ and $\nu$ 
of the metric perturbations, $h_{\mu\nu}^{(1)}$ and $h_{\mu\nu}^{(2)}$, 
and ignored $\Or((\mu/M)^3)$ and higher order terms. 
$G_{\mu\nu}^{(1)}$ is the well known linearized Hilbert-Einstein tensor, 
\begin{eqnarray}
\fl
G_{\mu\nu}^{(1)}[H] &=
-\frac{1}{2}H_{\mu\nu;\alpha}{}^{;\alpha}+H_{\alpha(\mu;\nu)}{}^{;\alpha}
-R_{\alpha\mu\beta\nu}H^{\alpha\beta} 
-\frac{1}{2}H_{\alpha}{}^{\alpha}{}_{;\mu\nu} -\frac{1}{2} g_{\mu\nu}
(H_{\lambda\alpha}{}^{;\alpha\lambda}
-H_{\alpha}{}^{\alpha}{}_{;\lambda}{}^{;\lambda}) \,, 
\end{eqnarray}
Here, $H_{\mu\nu}$ denotes $h_{\mu\nu}^{(1)}$ or $h_{\mu\nu}^{(2)}$, 
and semicolon ";" in the index 
denotes the covariant derivative 
with respect to the background metric. 
$G_{\mu\nu}^{(2)}$ consists of quadratic terms 
in the first order perturbations, 
\begin{eqnarray}
\fl
G_{\mu\nu}^{(2)}[h^{(1)},h^{(1)}] =
R_{\mu\nu}^{(2)}[h^{(1)},h^{(1)}] 
- \frac{1}{2}g_{\mu\nu}R_{\alpha}^{(2)}{}^{\alpha}[h^{(1)},h^{(1)}] 
\,; \\
\fl 
R_{\mu\nu}^{(2)}[h^{(1)},h^{(1)}] 
= \frac{1}{4}h^{(1)}_{\alpha\beta;\mu}h^{(1)}{}^{\alpha\beta}{}_{;\nu}
+\frac{1}{2}h^{(1)}{}^{\alpha\beta}(h^{(1)}_{\alpha\beta;\mu\nu}
+h^{(1)}_{\mu\nu;\alpha\beta}
-2h^{(1)}_{\alpha(\mu;\nu)\beta}) 
\nonumber \\ 
-\frac{1}{2}(h^{(1)}{}^{\alpha\beta}{}_{;\beta}-\frac{1}{2}h^{(1)}_{\beta}{}^{\beta;\alpha})
(2h^{(1)}_{\alpha(\mu;\nu)}-h^{(1)}_{\mu\nu;\alpha})
+\frac{1}{2}h^{(1)}_{\mu\alpha;\beta}h^{(1)}_{\nu}{}^{\alpha;\beta} 
-\frac{1}{2}h^{(1)}_{\mu\alpha;\beta}h^{(1)}_{\nu}{}^{\beta;\alpha} \,.
\nonumber 
\end{eqnarray}

On the other hand, the stress-energy tensor includes 
three parts; 
\begin{eqnarray}
T_{\mu\nu} &=
T_{\mu\nu}^{(1)}+T_{\mu\nu}^{(2,SF)}+T_{\mu\nu}^{(2,h)} 
\label{eq:T-exp} \,, 
\end{eqnarray}
The first order stress-energy tensor $T_{\mu\nu}^{(1)}$ 
which is the one of a point particle moving along a background geodesic, 
is given by 
\begin{eqnarray}
T^{(1)}{}^{\mu\nu} &= \mu \int^{+\infty}_{-\infty} \delta^{(4)}(x-z(\tau))
{dz^{\mu} \over d\tau}{dz^{\nu} \over d\tau}d\tau \,, 
\label{eq:pp}
\end{eqnarray}
where 
\begin{eqnarray}
z^{\mu} &=\{T(\tau),R(\tau),\Theta(\tau),\Phi(\tau)\} \,,
\end{eqnarray}
for the particle's orbit. 
$T_{\mu\nu}^{(2,SF)}$ denotes the deviation from the geodesic 
by the self-force as derived by the MiSaTaQuWa 
formalism~\cite{Mino:1996nk,Quinn:1996am}. 
We do not treat this stress-energy tensor explicitly in this paper. 
And $T_{\mu\nu}^{(2,h)}$, which is purely affected by 
the first order metric perturbations, is written as 
\begin{eqnarray}
T_{\mu\nu}^{(2,h)} = - \frac{1}{2}\, \mu \int^{+\infty}_{-\infty} 
h_{\alpha}^{(1)\alpha} \,\delta^{(4)}(x-z(\tau))
{dz^{\mu} \over d\tau}{dz^{\nu} \over d\tau}d\tau \,,
\label{eq:T2h}
\end{eqnarray}
where we have used the determinant 
\begin{eqnarray}
\tilde g&=g(1+h_{\alpha}^{(1)\alpha}) \,,
\end{eqnarray}
up to the first perturbative order. 

With the above expansion of the Hilbert-Einstein 
and stress-energy tensors, 
we may solve the following equation for the first perturbative order, 
\begin{eqnarray}
G_{\mu\nu}^{(1)}[h^{(1)}] &=&  8 \, \pi \,T_{\mu\nu}^{(1)} \,.
\label{eq:formal1st}
\end{eqnarray}
And for the second perturbative order, we have the following equation. 
\begin{eqnarray}
G_{\mu\nu}^{(1)}[h^{(2)}] &=& 
8 \, \pi \, \left( T_{\mu\nu}^{(2,SF)}+T_{\mu\nu}^{(2,h)} \right) 
- G_{\mu\nu}^{(2)}[h^{(1)},h^{(1)}] \,.
\label{eq:formal2nd}
\end{eqnarray}
Once the first order metric perturbations 
$h^{(1)}$ (and the self-force) are obtained, we may solve 
\eref{eq:formal2nd} with a second order source that can be considered as 
an effective stress-energy tensor. 
Systematically expanding the Hilbert-Einstein equations, one can obtain 
the perturbative equations order 
by order~\cite{Bruni:1996im,Campanelli:1998jv,Nakamura:2003wk,Brizuela:2007zz,Brizuela:2007zza}. 

%%%%%%%%%%%%%%%%%%%%%%%%%%%%%%%%%%%%%%%%%%%%%%%%%%%%%%%%%%%%%%%%%%%%%%
\section{First order perturbations in the Regge-Wheeler gauge} \label{sec:1st}
%%%%%%%%%%%%%%%%%%%%%%%%%%%%%%%%%%%%%%%%%%%%%%%%%%%%%%%%%%%%%%%%%%%%%%

In this paper, we consider the Schwarzschild background, 
\begin{eqnarray}
ds^2 &=
-\left(1-\frac{2M}{r}\right)dt^2
+\left(1-\frac{2M}{r}\right)^{-1} dr^2
+r^2\left(d\theta^2+\sin^2\theta d\phi^2\right) \,. 
\end{eqnarray}
in Boyer-Lindquist coordinates. 

Before considering the second perturbative order, 
it is necessary to discuss the first order metric perturbations, 
i.e., the first order Hilbert-Einstein equation given 
in \eref{eq:formal1st}. 
The Regge-Wheeler-Zerilli formalism~\cite{Regge:1957td,Zerilli:1971wd} 
is used here. 
The basic formalism has been given in Zerilli's paper~\cite{Zerilli:1971wd}, 
and it has been summarized in the time domain in~\cite{Lousto:2005ip,Lousto:2005xu}.
In \ref{app:RWZ}, we establish our notation and summarize 
the Regge-Wheeler-Zerilli formalism in the time domain. 

The treatment of the first perturbative order is as follows. 
First, we expand $h_{\mu\nu}^{(1)}$ and $T_{\mu\nu}^{(1)}$ 
in ten tensor harmonics components, given in \eref{eq:hharm} and 
\eref{eq:Tharm}. 
We then obtain the linearized field equations for each harmonic mode. 
Here, for example, for the even parity modes 
which have the even parity behavior, $(-1)^{\ell}$ 
under the transformation $(\theta,\phi)\to(\pi-\theta,\phi+\pi)$, 
we may consider the Zerilli equation in \eref{eq:zerilli-form}. 
Finally, imposing the Regge-Wheeler (RW) gauge conditions: 
\begin{eqnarray}
h_{0\,\ell m}^{(e)(1)RW}=h_{1\,\ell m}^{(e)(1)RW}=G_{\ell m}^{(1)RW}=0 \,, 
\end{eqnarray}
where the suffix ${\rm RW}$ stands for the RW gauge, 
we obtain the first order metric perturbations 
as in \eref{eq:recE}. 

%%%%%%%%%%%%%%%%%%%%%%%%%%%%%%
\subsection{Geodesic motion and the first order stress-energy tensor} 
%%%%%%%%%%%%%%%%%%%%%%%%%%%%%%

We consider a particle falling radially 
into a Schwarzschild black hole as the first order source. 
Assuming $\Theta(\tau)=\Phi(\tau)=0$, 
the equation of motion of the particle is given as 
\begin{eqnarray}
\left(\frac{dR(t)}{dt}\right)^2 &= -\left(1-\frac{2M}{R(t)}\right)^3 \frac{1}{E^2}
+ \left(1-\frac{2M}{R(t)}\right)^2 \,,
\end{eqnarray}
where $R(t)$ is the location of the particle 
and the energy $E$ is written by 
\begin{eqnarray}
E &= 
\left(1-\frac{2M}{R(t)} \right)\,\frac{dT(\tau)}{d\tau} \,.
\end{eqnarray}
We will also use 
\begin{eqnarray}
\frac{d^2R(t)}{dt^2} &= 
-\frac{3}{E^2}\left(1-\frac{2M}{R(t)}\right)^2 \frac{M}{R(t)^2}
+ 2\left(1-\frac{2M}{R(t)}\right) \frac{M}{R(t)^2} \,, 
\end{eqnarray}
to simplify equations. 
The tensor harmonics coefficients of the first order 
stress-energy tensor which are given in \tref{table:SET}, 
\begin{eqnarray}
{\cal A}^{(1)}_{\ell m}(t,r)
&=
\mu \,\displaystyle{\frac{E\,R(t)}{R(t)-{2M}}}\, 
\left(\frac{dR(t)}{dt}\right)^2 \frac{1}{(r-2M)^2}
\delta(r-R(t))Y_{\ell m}^*\left(0,0\right) \,, 
\nonumber \\ 
{\cal A}^{(1)}_{0\,\ell m}(t,r)
&=\mu \,\displaystyle{\frac{E\,R(t)}{R(t)-{2M}}}\, \frac{(r-2M)^2}{r^4} 
\delta(r-R(t))Y_{\ell m}^*\left(0,0\right) \,,
\nonumber \\ 
{\cal A}^{(1)}_{1\,\ell m}(t,r)
&=\sqrt{2}i\mu \,\displaystyle{\frac{E\,R(t)}{R(t)-{2M}}}\, 
\frac{dR(t)}{dt} \frac{1}{r^2}
\delta(r-R(t))Y_{\ell m}^*\left(0,0\right) \,.
\label{eq:source}
\end{eqnarray}
The remaining coefficients are zero. 
Because of the symmetry of the problem, 
we have only to consider the even parity modes.

%%%%%%%%%%%
\subsection{First order metric perturbations ($\ell \geq 2$)}
%%%%%%%%%%%

We introduce the following wave-function for the even parity 
($\ell \geq 2$) modes, 
\begin{eqnarray}
\fl
\psi^{\rm even}_{\ell m}(t,r) = 
\frac{2\,r}{\ell(\ell+1)} \biggl[
K_{\ell m}^{(1)RW} (t,r)
\nonumber \\ 
+2\,{\frac { ( r-2\,M )  }
{ ( r{\ell}^{2}+r\ell-2\,r+6\,M ) }}
\left(H_{2\,\ell m}^{(1)RW} ( t,r )
-r\,
{\frac {\partial }{\partial r}}K_{\ell m}^{(1)RW}(t,r)\right)
\biggr]
\,.
\label{eq:defpsi}
\end{eqnarray}
This function $\psi^{\rm even}_{\ell m}$ obeys the Zerilli equation, 
\begin{eqnarray}
\left(
-\frac{\partial^2}{{\partial t}^2} 
+ \frac{\partial^2}{{\partial r^*}^2}-V_\ell^{\rm even}(r)
\right)  \,\psi^{\rm even}_{\ell m}(t,r) 
= S^{\rm even}_{\ell m}(t,r) \,, 
\label{eq:zerilli}
\end{eqnarray}
where $r^*=r+2M \log (r/2M-1)$, the potential $V_{\ell}^{{\rm even}}$ 
is given in \eref{eq:Veven}. 
The source $S_{\ell m}^{\rm even}$ is calculated as in \eref{eq:Seven}.
For the $\ell=0$ and $1$ modes, 
we will need a different treatment as described in \sref{sec:firstL0}. 
(See also \cite{Sago:2002fe,Detweiler:2003ci}.)

The reconstruction of the first order metric perturbations 
from this wave-function in the RW gauge have been given 
in \eref{eq:recE}. 
In the head on collision case, the metric perturbations in 
the RW gauge are $C^0$ (continuous across the particle). 
One can see this as follows. 
First, using the following linearized Hilbert-Einstein equations 
for each harmonics mode, 
\begin{eqnarray}
\frac{H^{(1)}_{0\,\ell m}-H^{(1)}_{2\,\ell m}}{2} =
\frac{8\pi r^2 {\cal F}^{(1)}_{\ell m}}{\sqrt{\ell(\ell+1)(\ell-1)(\ell+2)/2}} 
\label{eq:G34} \,,
\end{eqnarray}
and ${\cal F}^{(1)}_{\ell m}=0$, 
we obtain $H^{(1)RW}_{2\,\ell m}=H^{(1)RW}_{0\,\ell m}$.
Then, removing $H^{(1)RW}_{1\,\ell m}$ 
from two linearized Hilbert-Einstein equations , 
\begin{eqnarray}
\fl
\frac{\partial}{\partial r}\left[
\left(1-\frac{2M}{r}\right)H^{(1)RW}_{1\,\ell m}\right]
-\frac{\partial}{\partial t}(H^{(1)RW}_{2\,\ell m}+K^{(1)RW}_{\ell m}) =
\frac{8\pi ir}{\sqrt{\ell(\ell+1)/2}}
{\cal B}^{(1)}_{0\,\ell m} \,, 
\label{eq:G13} \\
\fl
-\frac{\partial H^{(1)RW}_{1\,\ell m}}{\partial t}
+\left(1-\frac{2M}{r}\right)
\frac{\partial}{\partial r}(H^{(1)RW}_{0\,\ell m}-K^{(1)RW}_{\ell m})
+\frac{2M}{r^2}H^{(1)RW}_{0\,\ell m}
\nonumber \\ \hspace{-18mm}
+\frac{1}{r}\left(1-\frac{M}{r}\right)(H^{(1)RW}_{2\,\ell m}-H^{(1)RW}_{0\,\ell m})
 =
\frac{8\pi(r-2M)}{\sqrt{\ell(\ell+1)/2}}{\cal B}^{(1)}_{\ell m} \,, 
\label{eq:G23} 
\end{eqnarray}
we obtain the relation 
\begin{eqnarray}
\fl
\left[ -\frac{\partial^2}{\partial t^2} + 
\left(1-\frac{2M}{r}\right)^2 \frac{\partial^2}{\partial r^2} 
\right] H^{(1)RW}_{2\,\ell m}(t,r) = 
\left[ \frac{\partial^2}{\partial t^2} + 
\left(1-\frac{2M}{r}\right)^2 \frac{\partial^2}{\partial r^2} 
\right] K^{(1)RW}_{\ell m}(t,r)
\nonumber \\  + ({\rm 1st \,\, differential \,\, terms 
\,\, of \,\,} H^{(1)RW}_{2\,\ell m} \,\,{\rm and} \,\, K^{(1)RW}_{\ell m}) \,,
\end{eqnarray}
where we have used ${\cal B}^{(1)}_{0\,\ell m}={\cal B}^{(1)}_{\ell m}=0$. 
Therefore, we find 
that $H^{(1)RW}_{2\,\ell m}$ 
and $K^{(1)RW}_{\ell m}$ have 
the same differential behavior. 
Here, we note that the wave-function $\psi^{\rm even}_{\ell m}$ 
behaves as a step function around the particle location 
because of \eref{eq:zerilli} (and \eref{eq:explicitZ}). 
Thus, it is found that $\partial_r K^{(1)RW}_{\ell m} \sim \theta(r-R(t))$ 
with use of \eref{eq:defpsi}. 
This means that $K^{(1)RW}_{\ell m}$ is $C^0$. 
From \eref{eq:G23} with the above fact, 
$\partial_r H^{(1)RW}_{1\,\ell m} \sim \theta(r-R(t))$ 
is derived , i.e. $H^{(1)RW}_{1\,\ell m}$ is also $C^0$.
(See \cite{Lousto:1999za}.) 
We note that we can take up to second derivatives of 
the function $\psi^{\rm even}_{\ell m}$ 
with respect to $t$ and $r$ around the particle location 
as in \eref{eq:localQ}. 

In the next subsection, we treat only the $\ell=2$ mode 
which is the leading contribution in the first order perturbations. 
In this $\ell=2$ mode, we may consider only 
the $m=0$ mode because of $Y_{\ell m}\left(0,0\right)=0$ 
for $m \neq 0$ in \eref{eq:source}. 

%%%%%%%%%%%
\subsection{$\ell=2,\,m=0$ mode}
%%%%%%%%%%%

We focus here only on the $\ell=2,\,m=0$ mode.
For this mode, the Zerilli equation in \eref{eq:zerilli} becomes 
\begin{eqnarray}
\fl
\left[
-{\frac {\partial ^{2}}{\partial {t}^{2}}}
+\frac{\partial^2}{{\partial r^*}^2} 
-6\,{\frac {( r-2\,M) 
( 4\,{r}^{3} +4\,{r}^{2}M+6\,r{M}^{2}+3\,{M}^{3}) }{{r}^{4} ( 2\,r+3\,M ) ^{2}}}
\right]\,\psi^{\rm even}_{20}(t,r) 
\nonumber \\ \hspace{-18mm}
=-8\,\pi\,{\frac { \mu \, 
\left( 2\,{R(t)}^{2}-2\,R(t){E}^{2}M+6\,R(t)M+{M}^{2} \right) 
\left( R(t)-2\,M \right)^2  }{E {r}^{3}\left( 2\,r+3\,M \right) ^{2}}}
\nonumber \\  \hspace{-15mm} \times 
Y_{2 0}^*\left(0,0\right) 
\delta \left( r-R(t)  \right)
+\frac{8\,\pi }{3}\,{\frac { \mu\left( R(t)-2\,M \right) ^{3}
 }{E{R(t)}^{2} \left( 2\,R(t)+3\,M \right) }}
Y_{2 0}^*\left(0,0\right)\frac{d }{dr}\delta \left( r-R(t)  \right) \,,
\label{eq:explicitZ}
\end{eqnarray}
where we have used the formula 
\begin{eqnarray}
F(r)\frac{d}{dr} \delta'(r-R)=F(R) \frac{d}{dr}\delta(r-R)
- \frac{d}{dr} F'(r)\Big|_{r=R} \delta(r-R) \,,
\end{eqnarray} 
to simplify the source term. 

Here, we decompose the wave-function in the following form, 
\begin{eqnarray}
\psi^{\rm even}_{20}(t,r) &=  
\Psi_{20}^{out}(t,r)\theta(r-R(t)) + \Psi_{20}^{in}(t,r)\theta(R(t)-r)
\nonumber \\ 
&= \Psi_{20}^{\Theta}(t,r)\theta(r-R(t)) + \Psi_{20}^{H}(t,r) \,;
\nonumber \\ 
\Psi_{20}^{\Theta}(t,r) &= \Psi_{20}^{out}(t,r) - \Psi_{20}^{in}(t,r) \,,
\quad \Psi_{20}^{H}(t,r) = \Psi_{20}^{in}(t,r) \,,
\end{eqnarray}
where $\Psi_{20}^{out}$ and $\Psi_{20}^{in}$, 
i.e., also $\Psi^{\Theta}$ and $\Psi^{H}$, are homogeneous solution to 
the Zerilli equation. 
Using the fact that the first order metric perturbations are $C^0$, 
the following six quantities can be derived 
\begin{eqnarray}
\fl
\Psi_{20}^{\Theta}(t,r)|_{r=R(t)} = 
\frac{8\pi }{3}\,{\frac {\mu\,E\,R (t) }{2\,R (t) 
+3\,M}}Y_{20}^{*}\left(0,0\right) \,,
\nonumber \\ 
\fl
\frac{\partial}{\partial r} \Psi_{20}^{\Theta}(t,r)|_{r=R(t)} = 
16\pi\,{\frac {\mu\,E \, \left( R (t)^{2}
+R (t) M+{M}^{2} \right) }{ \left( 2\,R (t) +3\,M \right) ^{2} 
\left( 2\,M-R (t)  \right) }}Y_{20}^{*}\left(0,0\right) \,,
\nonumber \\ 
\fl
\frac{\partial}{\partial t} \Psi_{20}^{\Theta}(t,r)|_{r=R(t)} = 
-8\pi\,{\frac {\mu\,E\,R (t)  }
{ \left( 2\,R (t) +3\,M \right)  \left( 2\,M-R (t)  \right) }}\,\frac{dR(t)}{dt} 
Y_{20}^{*}\left(0,0\right) \,,
\nonumber \\ 
\fl
\frac{\partial^2}{\partial r^2} \Psi_{20}^{\Theta}(t,r)|_{r=R(t)} = 
-8\pi\,{\frac {\mu \,E\left( 7\,{M}^{3}-4\,R (t) {M}^{2}
+4\, R (t)^{2}M-8\, R (t)^{3} \right) 
 }{ \left( 2\,R (t) +3\,M \right) ^{3} 
\left( 2\,M-R (t)  \right) ^{2}}}Y_{20}^{*}\left(0,0\right) \,,
\nonumber \\ 
\fl
\frac{\partial^2}{\partial r \partial t} \Psi_{20}^{\Theta}(t,r)|_{r=R(t)} = 
8\pi\,{\frac {\mu\,E 
\left( -2\, R (t)^{2}+6\,R (t) M+3\,{M}^{2} \right) }
{ \left( 2\,R (t) +3\,M \right) ^{2} \left( 2\,M-R (t)  \right) ^{2}}}\,\frac{dR(t)}{dt} 
Y_{20}^{*}\left(0,0\right) \,,
\nonumber \\ 
\fl
\frac{\partial^2}{\partial t^2} \Psi_{20}^{\Theta}(t,r)|_{r=R(t)} = 
-8\pi\,{\frac {\mu \,E  M}{ \left( 2\,R (t) +3\,M \right) ^{3} 
R (t)^{2}}}Y_{20}^{*}\left(0,0\right) 
 \,.
\label{eq:localQ}
\end{eqnarray}
In the above equations, first we take the derivatives, 
and then set $r=R(t)$.
These quantities allow us to calculate 
the coefficients of the $\delta$ terms in the second order source.

%%%%%%%%%%%
\subsection{$\ell=0$ mode (Monopole perturbation)}\label{sec:firstL0}
%%%%%%%%%%%

Next, we consider the $\ell=0$ perturbation ($\ell=1$ modes can be 
completely eliminated in the center of mass coordinate system) 
which is present only in even parity. 
The metric perturbations and the gauge
transformation are given as 
\begin{eqnarray}
\bm{h}^{(1)}_{00} &= \left(1-\frac{2M}{r}\right)H^{(1)}_{0\,00}(t,r)\bm{a}_{0\,00}
-i \sqrt{2}H^{(1)}_{1\,00}(t,r)\bm{a}_{1\,00}
\nonumber \\ & 
+\left(1-\frac{2M}{r}\right)^{-1}H^{(1)}_{2\,00}(t,r)\bm{a}_{00}
+\sqrt{2}K^{(1)}_{00}(t,r)\bm{g}_{00}, \\
\xi^{(1)\mu}_{\ell=0} &=
\{V_0^{(1)}(t,r)Y_{00}(\theta,\phi),V_1^{(1)}(t,r)Y_{00}(\theta,\phi),0,0\} \,,
\label{eq:ggt0}
\end{eqnarray}
respectively.
The metric perturbations transforms under the above gauge 
transformation from the $G$ gauge to the $G'$ gauge as 
\begin{eqnarray}
\fl 
H_{0\,00}^{(1)G'}(t,r) = H_{0\,00}^{(1)G}(t,r)
+2\frac{\partial}{\partial t}V_0^{(1)G \to G'}(t,r)
+{2\,M \over r(r-2\,M)}V_1^{(1)G \to G'}(t,r) \,, 
\label{eq:gt00}
\\
\fl
H_{1\,00}^{(1)G'}(t,r) = H_{1\,00}^{(1)G}(t,r) 
- \frac{r}{r-2\,M}\frac{\partial}{\partial t}V_1^{(1)G \to G'}(t,r) 
+\frac{r-2\,M}{r}\frac{\partial}{\partial r} V_0^{(1)G \to G'}(t, r) \,, 
\label{eq:gt01}
\\
\fl
H_{2\,00}^{(1)G'}(t,r) = H_{2\,00}^{(1)G}(t,r) 
-2 \frac{\partial}{\partial r} V_1^{(1)G \to G'}(t,r) 
+ \frac{2\,M}{r(r-2\,M)} V_1^{(1)G \to G'}(r) 
\,, 
\label{eq:gt02}
\\
\fl
K_{00}^{(1)G'}(t,r) = K_{00}^{(1)G}(t,r) 
- \frac{2}{r} V_1^{(1)G \to G'}(t,r) \,.
\label{eq:gt03}
\end{eqnarray}

Here, we can choose $V_0^{(1)G \to G'}$ and $V_1^{(1)G \to G'}$ so that 
$H_{1\,00}^{(1)Z}=K_{00}^{(1)Z}=0$ where the suffix $Z$ stands 
for the Zerilli gauge~\cite{Zerilli:1971wd}. 
In this gauge, the two independent field equations are given by 
\begin{eqnarray}
\frac{\partial H_{2\,00}^{(1)Z}(t,r)}{\partial r}
+\frac{1}{r-2M}H_{2\,00}^{(1)Z}(t,r)
=\frac{8\pi r^3}{(r-2M)^2}{\cal A}_{0\,00}^{(1)}(t,r) \,, \\
\frac{\partial H_{0\,00}^{(1)Z}(t,r)}{\partial r}
+\frac{1}{r-2M}H_{2\,00}^{(1)Z}(t,r)
=-8\pi r {\cal A}_{00}^{(1)}(t,r) \,,
\end{eqnarray}
where ${\cal A}_{0\,00}^{(1)}$ and ${\cal A}_{00}^{(1)}$ 
are given in \eref{eq:source}. 
We solve the first equation, and then obtain 
\begin{eqnarray}
H_{2\,00}^{(1)Z}(t,r) = 8\pi\mu E \frac{1}{r-2M} Y_{00}^{*}(0,0) \,
\theta(r-R(t)) \,.
\end{eqnarray}
Next, substituting the above quantity into the second equation, 
we obtain
\begin{eqnarray}
\fl
H_{0\,00}^{(1)Z}(t,r) = 8\pi\mu E 
\left(\frac{1}{r-2M} - \frac{1}{R(t)-2M} 
- \frac{R(t)^2}{(R(t)-2M)^3}\left(\frac{dR(t)}{dt}\right)^2 \right)
\nonumber \\ \times
Y_{00}^{*}(0,0) \,
\theta(r-R(t)) \,.
\end{eqnarray}
It is difficult however to construct the second order source 
from the above metric perturbations, 
since these are not $C^0$. 
We instead consider a new (singular) gauge transformation, 
chosen to make the metric perturbations $C^0$. 

We consider the following gauge transformation. 
We call this the $C$ gauge, where 
the first order metric perturbations is $C^0$ at the particle location
\begin{eqnarray}
\fl
V_0^{(1)Z \to C}(t,r) =
\frac{2 \,\pi \,\mu \,Y_{00}^{*}(0,\,0)}{3\,E}\, 
\frac{\left( r-2\,M \right) \left( r+2\,M \right)  \left( {r}^{2}+4\,{M}^{2} \right)}
{ {r}^{4}} {\rm INT}(t)
\nonumber \\ \hspace{-20mm}
- \frac{2\,\pi \,\mu\,E\,\,Y_{00}^{*}(0,\,0)}{3}
\frac{\left( r-R (t)  \right) R(t) ^{3} }
{ \left( R (t) -2\,M \right) ^{4}{r}^{4}(r-2\,M)} \,{\frac {dR (t) }{dt}}
( -{r}^{2} R(t) ^{3}
- R(t) ^{2}{r}^{3}-{r}^{4}R (t) 
\nonumber \\ \hspace{-18mm}
+10\,{r}^{2}M R(t) ^{2}
+10\,MR (t) {r}^{3}
+8\,M{r}^{4} +10\,rM R(t) ^{3}
-16\,r{M}^{2} R(t) ^{2}-16\,{M}^{2}R (t) {r}^{2}
\nonumber \\ \hspace{-18mm} \left. 
-16\,{M}^{2}{r}^{3}
- R(t) ^{4}r+10\, R(t) ^{5}-62\, R(t) ^{4}M
+72\, R(t) ^{3}{M}^{2} \right) \theta (r-R(t)) 
\,, \\
\fl
V_1^{(1)Z \to C}(t,r) =
4\,\pi \,\mu\,EY_{00}^{*}(0,\,0)
\,{\frac {(r-2\,M)\, \left( r-R (t)  \right)  R(t) ^{6} }
{{r}^{6} \left( R (t) -2\,M \right) ^{2}}}
\,\theta (r-R(t)) \,; 
\\
\fl
{\rm INT}(t) = 
\int \! \biggl[
\left( -54\,{M}^{2}
+39\,R (t) M-50\,M{E}^{2}R (t) 
+12\,{E}^{2} R(t) ^{2} \right. \nonumber \\ 
\left. 
-6\, R(t) ^{2}
+16\,{M}^{2}{E}^{2} \right)
/\left( R (t) -2\,M \right) ^{3} 
\biggr] {dt} 
\nonumber \\ \hspace{-11mm} =
6\, {\frac {\left( -1+2\,{E}^{2} \right) E}{\sqrt {1-E^2 }}}
\arctan \left\{ \left( {\frac {M}{1-{E}^{2}}}
-R (t)  \right) \left[ R (t)  
\left( {\frac {2\,M}{1-{E}^{2}}}-R (t)  \right) \right]^{-1/2} \right\} 
\nonumber \\ \hspace{-9mm}
+12\,{E}^{2}\ln  \Biggl\{  \Biggl( {\frac {4\,{M}^{2}}{1-{E}^{2}}}
+{\frac {2\,R (t) M}{1-{E}^{2}}}-4\,R (t) M
 \nonumber \\ \qquad 
+\frac{4\,M\,E}{ \sqrt {1-{E}^{2}}}
\left[{R (t)  \left( {\frac {2\,M}{1-{E}^{2}}}
-R (t)  \right) }\right]^{1/2} \Biggr)  
\left( M(R (t) -2\,M) \right) ^{-1} \Biggr\} 
\nonumber \\ \hspace{-9mm}
+ \frac{\left( 13\, R(t) ^{2}
+48\,{M}^{2}-56\,R (t) M \right) E
[{{E}^{2}R (t) -R (t) +2\,M}]^{1/2}\sqrt {R (t) }} 
{ \left( R (t) -2\,M \right) ^{3}} \,. 
\label{eq:INT}
\end{eqnarray}
We then obtain the $\ell=0$ mode of the metric perturbations 
in this C gauge by using \eref{eq:gt00}, \eref{eq:gt01}, 
\eref{eq:gt02} and \eref{eq:gt03}, 
\begin{eqnarray}
\fl
H_{0\,00}^{(1)C}(t,r)  =
\frac{4\,\pi \,\mu\,Y_{00}^{*}(0,\,0)}{3}\,
\frac{\left( r-2\,M \right)  \left( r+2\,M \right)  \left( {r}^{2}+4\,{M}^{2} \right) }
{E \left( R (t) -2\,M \right) ^{3}{r}^{4}}( -54\,{M}^{2}+39\,R (t) M
\nonumber \\ \hspace{-18mm} \quad
-50\,M{E}^{2}R (t) 
+12\,{E}^{2} R(t) ^{2}
-6\, R(t) ^{2}
+16\,{M}^{2}{E}^{2} ) 
\,\theta (R(t)-r)
\nonumber \\ \hspace{-18mm}
- \frac{4\,\pi \,\mu\,Y_{00}^{*}(0,\,0)}{3}\,
\frac{1}{\left( R (t) -2\,M \right) ^{2}{r}^{7} \left( r-2\,M \right) E}
( -792\, R(t) ^{3}{r}^{3}{M}^{3}+792\, R(t) ^{2}{M}^{3}{r}^{4}
\nonumber \\ \hspace{-18mm}\quad
+192\,R (t) {r}^{3}{M}^{5}-864\,{r}^{3}{M}^{6}
+6\, R(t) ^{7}Mr{E}^{2}-398\, R(t) ^{5}{r}^{3}M+256\,{r}^{3}{M}^{6}{E}^{2}
\nonumber \\ \hspace{-18mm} \left. \quad 
-128\,{r}^{4}{M}^{5}{E}^{2}-6\, R(t) ^{6}M{r}^{2}{E}^{2}
+308\, R(t) ^{5}M{r}^{3}{E}^{2}
-468\, R(t) ^{4}{r}^{3}{M}^{2}{E}^{2}
\right. \nonumber \\ \hspace{-18mm} \left. \quad 
-72\, R(t) ^{3}{r}^{3}{M}^{3}{E}^{2}
-144\, R(t) ^{2}{r}^{3}{M}^{4}{E}^{2}
-672\,R (t) {r}^{3}{M}^{5}{E}^{2}+336\,R (t) {r}^{4}{M}^{4}{E}^{2}
\right. \nonumber \\ \hspace{-18mm} \left. \quad 
+50\, R(t) ^{6}{r}^{3}
-44\, R(t) ^{5}{r}^{4}-24\,{r}^{7}{M}^{2}{E}^{2}-50\, R(t) ^{6}{E}^{2}{r}^{3}
+72\, R(t) ^{2}{r}^{4}{M}^{3}{E}^{2}
\right. \nonumber \\ \hspace{-18mm} \left. \quad 
+24\,R (t) {r}^{7}M{E}^{2}+476\, R(t) ^{3}{r}^{4}{M}^{2}{E}^{2}+12\, R(t) ^{6}r{M}^{2}{E}^{2}
-282\, R(t) ^{4}{r}^{4}M{E}^{2}
\right. \nonumber \\ \hspace{-18mm} \left. \quad 
-6\, R(t) ^{2}{r}^{7}{E}^{2}
-96\,R (t) {M}^{4}{r}^{4}
+359\, R(t) ^{4}M{r}^{4}
-968\, R(t) ^{3}{M}^{2}{r}^{4}+432\,{r}^{4}{M}^{5}
\right. \nonumber \\ \hspace{-18mm} \left. \quad 
+1022\, R(t) ^{4}{r}^{3}{M}^{2}
-12\, R(t) ^{7}{M}^{2}{E}^{2}
+44\, R(t) ^{5}{r}^{4}{E}^{2} \right) 
\theta (r-R(t))
\,,\nonumber \\ 
\fl
H_{1\,00}^{(1)C}(t,r)  =
\frac{128\,\pi \,\mu\,Y_{00}^{*}(0,\,0)}{3\,E}
\,\frac{{M}^{4} \left( r-2\,M \right)}{r^6} \,{\rm INT}(t) 
\nonumber \\ \hspace{-18mm}
+\frac{4\,\pi\,\mu\,E\, Y_{00}^{*}(0,\,0)}{3}
\frac{\left( r-R (t)  \right) \,R(t) ^{5}}
{{r}^{6} \left( r-2\,M \right)  \left( R (t) -2\,M \right) ^{4}}
\, {\frac {d}{dt}}R (t) ( -72\,{M}^{2}{r}^{2}-288\,R (t) {M}^{3}
\nonumber \\ \hspace{-18mm} \quad 
+68\,rR (t) {M}^{2}+60\,MR (t) {r}^{2}
-132\,M R(t) ^{2}r+144\,r{M}^{3}-40\, R(t) ^{3}M+25\, R(t) ^{3}r
\nonumber \\ \hspace{-18mm} \quad
+248\, R(t) ^{2}{M}^{2}
-12\, R(t) ^{2}{r}^{2} ) 
\,\theta (r-R(t))
\,,\nonumber \\
\fl
H_{2\,00}^{(1)C}(t,r)  =
8\,\pi \,\mu\,EY_{00}^{*}(0,\,0)
\frac{ r-R (t) }{ \left( r-2\,M \right) {r}^{7} \left( R (t) -2\,M \right) ^{2}}
( -21\, R(t) ^{6}rM
+22\, R(t) ^{6}{M}^{2}
\nonumber \\ \hspace{-18mm} \quad 
+5\, R(t) ^{6}{r}^{2}
-4\, R(t) ^{5}M{r}^{2}
+ R(t) ^{5}{r}^{3}+4\, R(t) ^{5}r{M}^{2}
+4\, R(t) ^{4}{M}^{2}{r}^{2}+ R(t) ^{4}{r}^{4}
\nonumber \\ \hspace{-18mm} \left. \quad
-4\, R(t) ^{4}M{r}^{3}
-4\, R(t) ^{3}{r}^{4}M+4\, R(t) ^{3}{r}^{3}{M}^{2}
+ R(t) ^{3}{r}^{5}-4\, R(t) ^{2}M{r}^{5}
\right. \nonumber \\ \hspace{-18mm} \left. \quad
+4\, R(t) ^{2}{r}^{4}{M}^{2}
+ R(t) ^{2}{r}^{6}+4\,R (t) {M}^{2}{r}^{5}-4\,R (t) M{r}^{6}
+4\,{M}^{2}{r}^{6} \right) 
\,\theta (r-R(t))
\,,\nonumber \\ 
\fl
K_{00}^{(1)C}(t,r)  =
- 8\,\pi \,\mu\,EY_{00}^{*}(0,\,0)
{\frac { \left( r-2\,M \right) \, R(t) ^{6} 
\left( r-R (t)  \right) }{{r}^{7} \left( R (t) -2\,M \right) ^{2}}}
\,\theta (r-R(t))
\,.
\label{eq:L0inC}
\end{eqnarray}
Note that all of the above metric perturbations 
are $C^0$ at the particle location and go to zero at $r=\infty$ and the horizon. 
Using the metric perturbations in the $C$ gauge, 
we have discussed the second order source in \cite{Nakano:2007hv}. 
But, with this method, we can not obtain the second order 
gravitational wave at infinity 
because this is not an asymptotic flat gauge. 

We therefore must consider another treatment. 
The metric perturbations in \eref{eq:L0inC} 
have the following asymptotic behavior for large $r$ 
\begin{eqnarray}
\fl 
H_{0\,00}^{(1)C}(t,r) = 
\left(
8\,{\frac {\,\mu\,\pi \,E}{r}}+16\,{\frac {\mu\,\pi 
\,EM}{{r}^{2}}}
\right)
Y_{00}^*(0,0) + \Or(r^{-3}) \,, \,
H_{1\,00}^{(1)C}(t,r) = \Or(r^{-4})
\,,
\nonumber \\ 
\fl
H_{2\,00}^{(1)C}(t,r) = 
\left(8\,{\frac {\mu\,\pi \,E}{r}}
+16\,{\frac {\mu\,\pi 
\,EM}{{r}^{2}}}
\right)
Y_{00}^*(0,0) + \Or(r^{-3})
\,, \,
K_{00}^{(1)C}(t,r) = \Or(r^{-5}) \,.
\end{eqnarray}
Hence, the metric up to the first perturbative order 
in the system becomes 
\begin{eqnarray}
\fl
ds^2 = -\left(1-\frac{2M}{r}\right)(1-H_{0\,00}^{(1)C}(t,r)Y_{00}(\theta,\phi))dt^2
+ H_{100}^{(1)C}(t,r)Y_{00}(\theta,\phi)dtdr
\nonumber \\ \hspace{-15mm}
+\left(1-\frac{2M}{r}\right)^{-1}(1+H_{2\,00}^{(1)C}(t,r)Y_{00}(\theta,\phi))dr^2
+r^2(1+K_{00}^{(1)C}(t,r) Y_{00}(\theta,\phi)) d\Omega^2 
\nonumber \\ \hspace{-18mm} \sim 
-\left(1-\frac{2M+2\mu E}{r}\right)dt^2 
+ \left(1-\frac{2M+2\mu E}{r}\right)^{-1}dr^2
+r^2 d\Omega^2 \,.
\end{eqnarray}
We thus find that this perturbation is related 
to the mass increase of the system. 

From the above analysis, we define the total mass 
$M_{tot}=M+\mu\,E$ 
as that of the system. This means that 
the particle's mass is absorbed in the background 
Schwarzschild mass. 
Then, the first order $\ell=0$ renormalized metric perturbations become 
\begin{eqnarray}
H_{0\,00}^{(1)N}(t,r) &= 
H_{0\,00}^{(1)C}(t,r) - H_{0\,00}^{(1)M}(t,r) \,,
\nonumber \\ 
H_{2\,00}^{(1)N}(t,r) &= 
H_{2\,00}^{(1)C}(t,r) - H_{2\,00}^{(1)M}(t,r) \,,
\nonumber \\ 
H_{1\,00}^{(1)N}(t,r) &= 
H_{1\,00}^{(1)C}(t,r) \,,
\quad 
K_{00}^{(1)N}(t,r) = 
K_{00}^{(1)C}(t,r) \,,
\label{eq:L=0inN}
\end{eqnarray}
where we have labeled this renormalized metric perturbations 
by $N$, and 
\begin{eqnarray}
H_{0\,00}^{(1)M}(t,r) &= H_{2\,00}^{(1)M}(t,r) = 
8\,{\frac {\,\mu\,\pi \,E}{r-2\,M}}Y_{00}^*(0,0) \,,
\end{eqnarray}
Now for the asymptotic behavior for large $r$, we have 
\begin{eqnarray}
H_{0\,00}^{(1)N}(t,r) &= \Or(r^{-4})
\,, \quad 
H_{2\,00}^{(1)N}(t,r) &= \Or(r^{-5})
\,.
\end{eqnarray}
Next, we will use the coefficients of the 
first order metric perturbations labeled by $N$ in \eref{eq:L=0inN} 
to derive the second order source. 

%%%%%%%%%%%%%%%%%%%%%%%%%%%%%%%%%%%%%%%%%%%%%%%%%%%%%%%%%%%%%%%%%%%%%%
\section{Second order perturbations in the Regge-Wheeler gauge}\label{sec:2nd}
%%%%%%%%%%%%%%%%%%%%%%%%%%%%%%%%%%%%%%%%%%%%%%%%%%%%%%%%%%%%%%%%%%%%%%

%%%%%%%%%%%
\subsection{Second order Zerilli equation}
%%%%%%%%%%%

Since the first order metric perturbations 
contain only $m=0$ even parity modes, we can discuss the second order 
metric perturbations via the Zerilli equation. 
And we will choose the RW gauge condition. 
Here, we use a wave-function for the second perturbative order, 
\begin{eqnarray}
\chi_{20}^{\rm Z}(t,r)
=
\frac{1}{2\,r+3\,M}
\left({r}^2{\frac {\partial }{\partial t}}K_{20}^{(2)RW} ( t,r )
-( r-2\,M ) H_{1\,20}^{(2)RW} ( t,r ) \right) \,.
\end{eqnarray}
This is the same definition as in \eref{eq:ZerilliDef} 
for the first perturbative order. 
Here, we have considered the contribution 
from the $\ell=0$ and $2$ modes of the first perturbative order 
to the $\ell=2$ mode of the second perturbative order 
since this gives the leading contribution to gravitational radiation. 
This Zerilli function satisfies the equation, 
\begin{eqnarray}
\fl
\hat{\cal Z}_2^{\rm even} \,\chi_{20}^{\rm Z}(t,r)  = 
\left[
-{\frac {\partial ^{2}}{\partial {t}^{2}}}
+\frac{\partial^2}{{\partial r^*}^2}
-6\,{\frac {( r-2\,M) 
( 4\,{r}^{3} +4\,{r}^{2}M+6\,r{M}^{2}+3\,{M}^{3}) }{{r}^{4} ( 2\,r+3\,M ) ^{2}}}
\right] \,
\chi_{20}^{\rm Z}(t,r) 
\nonumber \\ 
= {\cal S}^{\rm{Z}}_{20}(t,r) \,;
\\
\fl
{\cal S}^{\rm{Z}}_{20}(t,r) = 
 {\frac {8\,\pi \,\sqrt {3} \left( r-2\,M \right) ^{2}}
{3(2\,r+3\,M)}}\,{\frac {\partial }{\partial t}}{\cal B}^{(2)}_{20} ( t,r ) 
+ {\frac { 8\,\pi \left( r-2\,M \right) ^{2}}{2\,r+3\,M}}
\,{\frac {\partial }{\partial t}} {\cal A}^{(2)}_{20} ( t,r ) 
\nonumber \\ \hspace{-8mm}
- \frac{8\,\sqrt {3}\, \pi \left( r-2\,M \right) }{3}
{\frac {\partial }{\partial t}}{\cal F}^{(2)}_{20} ( t,r ) 
- {\frac {4\,\sqrt {2}\,i\pi  \left( r-2\,M \right) ^{2} }
{2\,r+3\,M}}{\frac {\partial }{\partial r}}{\cal A}^{(2)}_{1\,20} ( t,r ) 
\nonumber \\ \hspace{-8mm}
- {\frac {8\,\sqrt {2}\,i\pi  \left( r-2\,M \right)  \left( 5\,r-3\,M \right) M}
{r \left( 2\,r+3\,M \right) ^{2}}}{\cal A}^{(2)}_{1\,20} ( t,r ) 
- {\frac {8\,\sqrt {3} \,i\pi \left( r-2\,M \right) ^{2}}{3(2\,r+3\,M)}}
{\frac {\partial }{\partial r}}{\cal B}^{(2)}_{0\,20} ( t,r ) 
\nonumber \\ \hspace{-8mm}
+ {\frac {32\,\sqrt {3}\,i\pi \,
\left( 3\,{M}^{2}+{r}^{2} \right)  \left( r-2\,M \right) }
{3\,r \left( 2\,r+3\,M \right) ^{2}}} {\cal B}^{(2)}_{0\,20} ( t,r ) 
\,.
\label{eq:2ndS22}
\end{eqnarray}
The functions ${\cal B}^{(2)}_{20}$ etc. are derived 
from the effective stress-energy tensor on the 
left hand side of \eref{eq:formal2nd}, 
\begin{eqnarray}
T_{\mu\nu}^{(2,eff)} &=& 
\left( T_{\mu\nu}^{(2,SF)}+T_{\mu\nu}^{(2,h)} \right) 
- \frac{1}{8 \, \pi}G_{\mu\nu}^{(2)}[h^{(1)},h^{(1)}] \,,
\end{eqnarray}
by the same tensor harmonics expansion as for the first perturbative order. 

The second order metric perturbations from $\chi_{20}^{\rm Z}$ 
in the RW gauge are given by 
\begin{eqnarray}
\fl 
{\frac {\partial }{\partial t}}K_{20}^{(2)RW} \left( t,r \right) = 
6\,{\frac { \left( {r}^{2}+rM+{M}^{2} \right) 
 }{ \left( 2
\,r+3\,M \right) {r}^{2}}}\chi_{20}^{\rm Z} \left( t,r \right)
+{\frac { \left( r-2\,M \right)  }{r}}{\frac {
\partial }{\partial r}}\chi_{20}^{\rm Z} \left( t,r \right)
\nonumber \\  \fl \qquad \qquad \qquad \quad 
+{\frac {4\,\sqrt {2}\,i\pi 
r \left( r-2\,M \right) }{2\,
r+3\,M}}{\cal A}^{(2)}_{1\,20} \left( t,r \right) 
+{\frac {8\sqrt {3}\,i\pi 
r \left( r-2\,M \right) }{3(2\,r+3\,M)}}{\cal B}^{(2)}_{0\,20} \left( t,r \right)
 \,,
\label{eq:2ndKRW}
\\ 
\fl
{\frac {\partial }{\partial t}}H_{2\,20}^{(2)RW} \left( t,r \right) 
= 
r{\frac {
\partial ^{2}}{\partial r\partial t}}K_{20}^{(2)RW} 
\left( t,r \right) +3\,{\frac {
M}{{r}^{2}}}\chi_{20}^{\rm Z} \left( t,r \right) 
-{\frac { \left( 2\,r+3\,M
 \right) }{r}} {\frac {\partial }{\partial r}}\chi_{20}^{\rm Z} \left( t,r \right)
\nonumber \\  \fl \qquad \qquad \qquad \quad 
-\frac{8\sqrt {3}}{3}\,i\pi \,{\cal B}^{(2)}_{0\,20} \left( t,r \right) r
 \,,
\\ 
\fl
H_{0\,20}^{(2)RW} \left( t,r \right) = 
H_{2\,20}^{(2)RW} \left( t,r \right) +\frac{8\sqrt {3}}{3}
\,\pi \,{\cal F}_{20}^{(2)} \left( t,r \right) {r}^{2} \,.
\end{eqnarray}
Since the RW gauge is not asymptotically flat, 
we need to derive the second order metric perturbations 
in an asymptotic flat (AF) gauge 
to obtain the second order gravitational wave at spatial infinity.
This is discussed in \ref{app:SOGT}. 

Note that in \eref{eq:T2h}. 
the delta function $\delta^{(4)}(x-z(\tau))$ 
in $T_{\mu\nu}^{(2,h)}$ includes an angular dependence, 
$
\delta^{(2)}(\Omega-\Omega(\tau))
=\sum_{\ell m}Y_{\ell m}(\Omega)Y_{\ell m}^*(\Omega(\tau)) \,. 
$
We have considered only the contribution from the $\ell=0$ and $2$ modes 
of the first order perturbations. Consistently, we must use 
only the three components with the factors, 
$h_{\alpha}^{(1)\alpha}(\ell=2)Y_{2m}(\Omega)Y_{2m}^*(\Omega(\tau))$, 
$h_{\alpha}^{(1)\alpha}(\ell=2)Y_{0m}(\Omega)Y_{0m}^*(\Omega(\tau))$
and $h_{\alpha}^{(1)\alpha}(\ell=0)Y_{2m}(\Omega)Y_{2m}^*(\Omega(\tau))$. 

In the second order source, given in \eref{eq:2ndS22}, 
we may wander if there is any $\delta^2$ term 
which prevents us from making the second order calculation.
The answer is ``No''. This is because, 
in the head-on collision case, 
the first order metric perturbations in the RW gauge are $C^0$: 
$G_{\mu\nu}^{(2)}[h^{(1)},h^{(1)}]$ includes 
second derivatives and 
we need one more derivative to construct the second order source 
of \eref{eq:2ndS22}. 
Here, $(h^{(1)})^2$ is $C^0 \times C^0$ and 
its third derivative yields $C^0 \times \delta'$ and 
$\theta \times \delta$ as the most singular terms. 
On the other hand, $T_{\mu\nu}^{(2,h)}$ includes 
only $C^0 \times \delta$ terms. 
Note also that there is no $\delta^2$ terms coming 
from $T_{\mu\nu}^{(2,SF)}$, (which we ignored otherwise in this paper.) 
Using the result of \cite{Lousto:1999za,Barack:2002ku}, we can include 
the contributions of $T_{\mu\nu}^{(2,SF)}$. 

In the following subsection, we derive
the second order source of the Zerilli equation in \eref{eq:2ndS22}. 
The summary is given here. 
From \eref{eq:2ndS22}, we obtain the second order source 
as 
\begin{eqnarray}
{\cal S}^{\rm{Z}}_{20}(t,r) &=& {}^{(2,2)}{\cal S}^{\rm{Z}}_{20}(t,r)
+{}^{(0,2)}{\cal S}^{\rm{Z}}_{20}(t,r) \,, 
\end{eqnarray}
where ${}^{(2,2)}{\cal S}^{\rm{Z}}_{20}$ and ${}^{(0,2)}{\cal S}^{\rm{Z}}_{20}$ 
are the contribution from $(\ell=2)\cdot(\ell=2)$ and 
$(\ell=0)\cdot(\ell=2)$, respectively. 
Note that while the above source term is locally integrable 
near the particle's location, some terms diverge as $r\to\infty$ or $2M$.
This is not readily suitable for numerical calculations. 
We then consider some regularization 
for the asymptotic behavior. (See e.g. \cite{Gleiser:1998rw}.) 
In order to obtain a second order source which behaves well everywhere, 
we define a regularized Zerilli function by 
\begin{eqnarray}
{\tilde \chi}_{20}^{\rm Z}(t,r) = \chi_{20}^{\rm Z}(t,r) 
- \chi_{20}^{{\rm reg},(2,2)} - \chi_{20}^{{\rm reg},(0,2)} \,. 
\label{eq:formalreg}
\end{eqnarray}
The best suited equation to solve the Zerilli equation numerically 
is then
\begin{eqnarray}
\hat{\cal Z}_2^{\rm even} 
{\tilde \chi}_{20}^{\rm Z}(t,r) &= {\cal S}^{\rm{Z},reg}_{20}(t,r) \,,
\label{eq:regZeq}
\end{eqnarray}
where the regular source ${\cal S}^{\rm{Z},reg}_{20}$ is given by 
\begin{eqnarray}
{\cal S}^{\rm{Z},reg}_{20}(t,r) &= 
\left({}^{(2,2)}{\cal S}^{\rm{Z}}_{20}(t,r) 
- \hat{\cal Z}_2^{\rm even} \chi_{20}^{{\rm reg},(2,2)}(t,r)\right)
\nonumber \\ & \quad 
+\left({}^{(0,2)}{\cal S}^{\rm{Z}}_{20}(t,r)
- \hat{\cal Z}_2^{\rm even} \chi_{20}^{{\rm reg},(0,2)}(t,r)\right) \,. 
\label{eq:regZeqS}
\end{eqnarray}

%%%%%%%%%%%
\subsection{Regularized second order source from $(\ell=2)\cdot(\ell=2)$}
%%%%%%%%%%%

When we consider the asymptotic behavior of the second order source 
for large $r$, we use the retarded solution of the first order 
wave-function $\psi_{\ell m}^{\rm even}$ 
with the retarded time $t-r^*$, which we expand in 
inverse powers of $r$. The wave-function becomes 
\begin{eqnarray}
\psi_{20}^{\rm even}(t,r) = F_I'(t-r^*)+\frac{3}{r} F_I(t-r^*) 
+ \Or(r^{-2}) \,,
\end{eqnarray}
where $F_I$ is some function of $(t-r_*)$ 
and $F_I'(x)=d F(x)/dx$. 

On the other hand, the wave-function is expanded 
near the horizon as, 
\begin{eqnarray}
\psi_{\ell m}^{\rm even}(t,r) &= F_H'(t+r^*)+\frac{1}{4M}F_H(t+r^*) 
+ \frac{27(r-2M)}{56M^2} F_H(t+r^*) 
\nonumber \\ & \quad 
+ \Or((r-2M)^2) \,.
\end{eqnarray}
Using these expansions , 
we derive a second order source which is regular everywhere. 

In order to obtain a well behaved source for large values of $r$, 
we define a regularization function by 
\begin{eqnarray}
\chi_{20}^{{\rm reg},(2,2)}(t,r) = \frac{\sqrt{5}}{7\,\sqrt{\pi}}
{\frac {{r}^{2}}{2\,r+3\,M}} 
\left( {\frac {\partial }{\partial t}}K_{20}^{(1)RW} ( t,r )  \right) 
K_{20}^{(1)RW} ( t,r ) \,.
\label{eq:Reg22}
\end{eqnarray}
Note that the regularization function is not unique. 
Therefore, this affects 
the formal expression of the second order gravitational wave as in \eref{eq:FINAL}.
However, for any specific computation, there is no ambiguity in the physical final results.

Using the above regularization function, 
the regularized second order source ${}^{(2,2)}{\cal S}^{\rm{Z},reg}_{20}$ 
from the $(\ell=2)\cdot(\ell=2)$ coupling is obtained 
\begin{eqnarray}
\fl
{}^{(2,2)}{\cal S}^{\rm{Z},reg}_{20}(t,r) = \left({}^{(2,2)}{\cal S}^{\rm{Z}}_{20}(t,r) 
- \hat{\cal Z}_2^{\rm even} \chi_{20}^{{\rm reg},(2,2)}(t,r)\right)
\nonumber \\ 
= {}^{G(2,2)}S^{out}_{20}(t,r)\theta(r-R(t))
+ {}^{G(2,2)}S^{in}_{20}(t,r)\theta(R(t)-r)
\nonumber \\ \quad
+ {}^{G(2,2)}S^{\delta}_{20}(t,r)\delta(r-R(t))
+ {}^{G(2,2)}S^{\delta'}_{20}(t,r)\frac{d}{dr}\delta(r-R(t))
\nonumber \\ \quad 
+ {}^{T(2,2)}S^{\delta}_{20}(t,r)\delta(r-R(t)) 
+ {}^{T(2,2)}S^{\delta'}_{20}(t,r)\frac{d}{dr}\delta(r-R(t)) \,,
\label{eq:S22Zreg}
\end{eqnarray}
where the superscripts $G$ and $T$ denote the source terms 
which are derived from $G_{\mu\nu}^{(2)}[h^{(1)},h^{(1)}]$ and 
$T_{\mu\nu}^{(2,h)}$, respectively. 
The six source factors, ${}^{G(2,2)}S^{out}_{20}$, etc. 
are given in \ref{app:sost}. 
For the factors of the step functions in the above equation, 
${}^{G(2,2)}S^{out}_{00}$ behaves as $\Or(r^{-2})$ for large $r$,  
and ${}^{G(2,2)}S^{in}_{20}$ vanishes as $\Or((r-2M)^1)$ at $r=2M$.

%%%%%%%%%%%
\subsection{Regularized second order source from $(\ell=0)\cdot(\ell=2)$}
%%%%%%%%%%%

The second order source derived from the $(\ell=0)\cdot(\ell=2)$ coupling 
needs regularization of its behavior near the horizon. 
Note that when we choose another gauge for the first order $\ell=0$ mode, 
some regularization is also needed for the behavior for large $r$. 
To that end, we use the regularization function, 
\begin{eqnarray}
\fl
\chi_{20}^{{\rm reg},(0,2)}(t,r) = 
{\frac {1}{5488\sqrt {\pi }}}\,\frac{ 1061\,r-2728\,M }{r}
\,  H_{2\,00}^{(1)N}(t,r)
 H_{2\,20}^{(1)RW} \left( t,r \right) 
\nonumber \\ 
-{\frac {107}{2744\sqrt {\pi }}}\,{\frac { \left( r-2\,M \right) M
}{r}}\,H_{2\,00}^{(1)N}(t,r)
 {\frac {\partial }{\partial r}}H_{2\,20}^{(1)RW} \left( t,r \right)
\nonumber \\ 
+{\frac {1}{2744\sqrt {\pi }}}
\,{\frac {M \left( 583\,r-756\,M
 \right) }{r}}H_{2\,00}^{(1)N}(t,r) {\frac {\partial }{\partial t}}
H_{2\,20}^{(1)RW} \left( t,r \right)  \,.
\label{eq:Reg20}
\end{eqnarray}

Then, we obtain the second order regularized source 
from the $(\ell=0)\cdot(\ell=2)$ coupling
\begin{eqnarray}
\fl
{}^{(0,2)}{\cal S}^{\rm{Z}}_{20}(t,r) = 
+\left({}^{(0,2)}{\cal S}^{\rm{Z}}_{20}(t,r)
- \hat{\cal Z}_2^{\rm even} \chi_{20}^{{\rm reg},(0,2)}(t,r)\right)
\nonumber \\
= {}^{G(0,2)}S^{out}_{00}(t,r)\theta(r-R(t))
+ {}^{G(0,2)}S^{in}_{20}(t,r)\theta(R(t)-r)
\nonumber \\ \quad
+ {}^{G(0,2)}S^{\delta}_{20}(t,r)\delta(r-R(t))
+ {}^{G(0,2)}S^{\delta'}_{20}(t,r)\frac{d}{dr}\delta(r-R(t))
\nonumber \\ \quad
+ {}^{T(0,2)}S^{\delta}_{20}(t,r)\delta(r-R(t)) 
+ {}^{T(0,2)}S^{\delta'}_{20}(t,r)\frac{d}{dr}\delta(r-R(t)) 
\nonumber \\ \quad
+ {}^{T(2,0)}S^{\delta}_{20}(t,r)\delta(r-R(t)) 
+ {}^{T(2,0)}S^{\delta'}_{20}(t,r)\frac{d}{dr}\delta(r-R(t)) \,. 
\label{eq:S02Zreg}
\end{eqnarray}
Here, we have written the contribution of 
$h_{\alpha}^{(1)\alpha}(\ell=2)Y_{00}(\Omega)Y_{00}^*(\Omega(\tau))$ and 
$h_{\alpha}^{(1)\alpha}(\ell=0)Y_{20}(\Omega)Y_{20}^*(\Omega(\tau))$ 
as ${}^{T(0,2)}S^{\delta}_{20}$ and ${}^{T(2,0)}S^{\delta}_{20}$, 
respectively. In the above equation, 
${}^{G(0,2)}S^{out}_{00}$ behaves as $\Or(r^{-2})$ for large $r$,  
and ${}^{G(0,2)}S^{in}_{20}$ vanishes as $\Or((r-2M)^1)$ at $r=2M$.

%%%%%%%%%%%%%%%%%%%%%%%%%%%%%%%%%%%%%%%%%%%%%%%%%%%%%%%%%%%%%%%%%%%%%%
\section{Summary and Discussion}\label{sec:dis} 
%%%%%%%%%%%%%%%%%%%%%%%%%%%%%%%%%%%%%%%%%%%%%%%%%%%%%%%%%%%%%%%%%%%%%%

In this paper, we have completed, for the first time the program of self-consistent
binary black hole second order perturbations in the small mass ratio limit. Our
analysis applies to headon collision, but can be extended to arbitrary orbits if
worked in the Lorenz gauge \cite{Barack:2005nr}. The first order perturbations
of two black holes starting from rest at a finite distance have been solved
in the frequency \cite{Lousto:1996sx} and time \cite{Lousto:1997wf} domains.
Then the corrected trajectories (from the background geodesics)
via the computation of the self-force have been computed in \cite{Lousto:1999za,Barack:2002ku}.
Here, we obtained the regularized source 
for the second order Zerilli equation in the case of a particle falling radially 
into a Schwarzschild black hole. This is given by \eref{eq:regZeqS}
with \eref{eq:S22Zreg} and \eref{eq:S02Zreg}. 
Using this second order source, 
we are able to compute the second order contribution to gravitational radiation
by numerical integration of the wave equation \eref{eq:regZeq} and compare to full
numerical simulations. 
The derivation of the second order gravitational wave from 
the second order Zerilli function is discussed in \ref{app:SOGT}.

A key point is to prove that there is no $\delta^2$ term in the second order source. 
To show this, we have used the fact that the first order metric perturbations 
in the RW gauge are $C^0$. 
In general orbit case (including circular orbits), 
the first order metric perturbations are not $C^0$ in the RW gauge, 
but they are $C^0$ in the Lorenz gauge~\cite{Barack:2005nr}.
This later gauge choice favors the study of 
the second order perturbations for generic orbits. 

To be fully consistent in the second perturbative order, 
we have to include the self-force contribution $T_{\mu\nu}^{(2,SF)}$ 
which is derived from the deviation from the background geodesic motion. 
The self-force for a head-on collision has been calculated in~\cite{Lousto:1999za,Barack:2002ku}, 
and in a circular orbit around a Schwarzschild black hole 
in~\cite{Nakano:2003he,Barack:2007tm,Detweiler:2008ft}, 
but have not been obtained in the general case yet. 

Here, we have discussed only the $\ell=2$ mode 
of the second perturbative order since this gives the leading contribution 
to gravitational radiation. There remain a question about 
the convergence of the second order metric perturbations, though. 
Based on the works 
by Rosenthal~\cite{Rosenthal:2005it,Rosenthal:2006nh, Rosenthal:2006iy}, 
we discuss this problem. 

First, we consider a second order wave equation 
which is given by 
\begin{eqnarray}
\Box h^{(2)} = S^{(2)}_h \,, 
\end{eqnarray}
where $\Box$ is the wave operator and 
the second order source $S^{(2)}_h$ is derived from 
the first order metric perturbations with a local behavior 
around the particle location as 
$
h^{(1)} \sim \Or(\epsilon^{-1}) \,,
$
where the spatial separation $\epsilon=|{\bf x}-{\bf x_0}|$ 
with a particle location ${\bf x_0}$. 
In this discussion, we are not on the particle's world line 
but rather take a limit to the particle location. 
We need second derivatives 
to compute $S^{(2)}_h$, then the local behavior becomes 
$
S^{(2)}_h \sim \Or(\epsilon^{-4}) \,.
$
For the above source, if we solve the wave equation 
by using a usual four dimensional Green's function method, 
this solution diverge everywhere~\cite{Rosenthal:2005it}. 
Therefore, to obtain the finite solution, 
we need to remove the $\Or(\epsilon^{-4})$ and $\Or(\epsilon^{-3})$ terms 
from $S^{(2)}_h$. Here, we note that we can remove the $\Or(\epsilon^{-3})$ terms 
by using a regular gauge transformation~\cite{Rosenthal:2006nh}. 
Since we consider the second order gravitational wave at infinity 
which is gauge invariant, the $\Or(\epsilon^{-3})$ terms does not contribute. 
In practice, we use the gauge invariant wave-function in our calculation. 
Thus, we may discuss only the problematic $\Or(\epsilon^{-4})$ terms. 

Here, Rosenthal has already shown the most singular part of 
the second order metric perturbations as 
\begin{eqnarray}
h^{(2)}_s \sim \Or(\epsilon^{-2}) \,.
\end{eqnarray}
This is a peculiar solution of 
$
\Box h^{(2)}_s \sim \Or(\epsilon^{-4}) \,.
$
Using this solution, we rewrite the second order wave solution as 
\begin{eqnarray}
h^{(2)}_r 
= h^{(2)}-h^{(2)}_s
\sim \Box^{-1} \left(S^{(2)}-\Or(\epsilon^{-4})\right) 
\sim \Box^{-1} \left(\Or(\epsilon^{-2})\right) \,,
\end{eqnarray}
The right hand side of the last line is finite after the integration 
by using the retarded Green's function as $\Box^{-1}$. 
The final result is, 
\begin{eqnarray}
h^{(2)}=h^{(2)}_s+h^{(2)}_r \,,
\end{eqnarray}
is the physical second-order gravitational perturbations~\cite{Rosenthal:2006nh}. 
Thus, if we construct a second order gauge invariant $\psi^{(2)}$
from the above metric perturbations, this is given by 
\begin{eqnarray}
\psi^{(2)} = \psi^{(2)}_s + \psi^{(2)}_r \,,
\label{eq:GIpsi2}
\end{eqnarray}
where $\psi^{(2)}_s$ and $\psi^{(2)}_r$ are 
derived from $h^{(2)}_s$ and $h^{(2)}_r$, respectively. 
As a result, the apparent divergence derived from 
the first consideration is only a gauge contribution. 

Next, we consider 
the expansion in terms of tensor harmonics of the second order gauge invariant wave-function 
(which needs not to be same as the Regge-Wheeler or Zerilli function), 
\begin{eqnarray}
\Box_{\ell m} \psi^{(2)}_{\ell m} = S^{(2)}_{\ell m} \,.
\end{eqnarray}
In our situation, we solve the above equation 
by numerical integration. Formally, we can write the solution as 
$
\psi^{(2)}_{\ell m} = \Box^{-1}_{\ell m} S^{(2)}_{\ell m} \,.
$
This $\Box^{-1}_{\ell m}$ means a numerical integration 
with an appropriate boundary condition. 
We may also use the retarded Green's function. 
Since the solution $\psi^{(2)}_{\ell m}$ is gauge invariant, 
this summation over the $(\ell,\,m)$ modes coincides with 
$\psi^{(2)}$ in \eref{eq:GIpsi2}. Hence, 
the summation of $\psi^{(2)}_{\ell m}$ over modes has a finite value, 
(except for the location of the particle.) 
In particular, the asymptotic behaviour for large $r$, where we need to
compute gravitational radiation, is well defined.

\ack

We would like to thank K.~Ioka, K.~Nakamura and T.~Tanaka for useful discussions. 
This work is supported by JSPS for Research Abroad (HN)
and by NSF 
grants PHY-0722315,  PHY-0701566, PHY-0714388, and PHY-0722703, and
from a NASA grant 07-ATFP07-0158.

\appendix

%%%%%%%%%%%%%%%%%%%%%%%%%%%%%%%%%%%%%%%%%%%%%%%%%%%%%%%%%%%%%%%%%%%%%%
\section{The Regge-Wheeler-Zerilli formalism in the time domain}\label{app:RWZ}
%%%%%%%%%%%%%%%%%%%%%%%%%%%%%%%%%%%%%%%%%%%%%%%%%%%%%%%%%%%%%%%%%%%%%%

In this paper, we have discussed the metric perturbations 
using the Regge-Wheeler-Zerilli formalism, 
with similar notation to that of Zerilli's paper~\cite{Zerilli:1971wd}. 
There are some differences though
that we summarize in this appendix. 

For the first and second order metric perturbations, 
and stress-energy tensors, 
we expand $h_{\mu\nu}^{(i)}$
and $T_{\mu\nu}^{(i)}$ ($i=1,\,2$) in tensor harmonics, 
\begin{eqnarray}
\fl 
\bm{h}^{(i)} = \sum_{\ell m} \left[
\left(1-\frac{2M}{r}\right)H^{(i)}_{0\,\ell m}(t,r)\bm{a}_{0\,\ell m}
-i\sqrt{2}H^{(i)}_{1\,\ell m}(t,r)\bm{a}_{1\,\ell m}
\right. \nonumber \\ \hspace{-12mm}
+\left(1-\frac{2M}{r}\right)^{-1}H^{(i)}_{2\,\ell m}(t,r)\bm{a}_{\ell m}
-\frac{i}{r}\sqrt{2\ell(\ell+1)}h^{(e)(i)}_{0\,\ell m}(t,r)\bm{b}_{0\,\ell m}
\nonumber \\ \hspace{-12mm}
+\frac{1}{r}\sqrt{2\ell(\ell+1)}h^{(e)(i)}_{1\,\ell m}(t,r)\bm{b}_{\ell m}
+\left[{\frac{1}{2}\ell(\ell+1)(\ell-1)(\ell+2)}\right]^{1/2}
G^{(i)}_{\ell m}(t,r)\bm{f}_{\ell m}
\nonumber \\ \hspace{-12mm}
+\left(\sqrt{2}K^{(i)}_{\ell m}(t,r)
 -\frac{\ell(\ell+1)}{{\sqrt{2}}}G^{(i)}_{\ell m}(t,r)\right)\bm{g}_{\ell m}
-\frac{\sqrt{2\ell(\ell+1)}}{r}h^{(i)}_{0\,\ell m}(t,r)\bm{c}_{0\,\ell m}
\nonumber \\ \left. \hspace{-12mm}
+\frac{i\sqrt{2\ell(\ell+1)}}{r}h^{(i)}_{1\,\ell m}(t,r)\bm{c}_{\ell m}
+\frac{\left[{2\ell(\ell+1)(\ell-1)(\ell+2)}\right]^{1/2}}{2r^2}
h^{(i)}_{2\,\ell m}(t,r)\bm{d}_{\ell m}
\right] \,, \label{eq:hharm}
\\
\fl
\bm{T}^{(i)} = \sum_{\ell m}\left[
{\cal A}^{(i)}_{0\,\ell m}\bm{a}_{0\,\ell m}
+{\cal A}^{(i)}_{1\,\ell m}\bm{a}_{1\,\ell m}
+{\cal A}^{(i)}_{\ell m}\bm{a}_{\ell m}
+{\cal B}^{(i)}_{0\,\ell m}\bm{b}_{0\,\ell m}
+{\cal B}^{(i)}_{\ell m}\bm{b}_{\ell m}
 \right. \nonumber \\  \hspace{-5mm} \left. 
+{\cal Q}^{(i)}_{0\,\ell m}\bm{c}_{0\,\ell m}
+{\cal Q}^{(i)}_{\ell m}\bm{c}_{\ell m}
+{\cal D}^{(i)}_{\ell m}\bm{d}_{\ell m}
+{\cal G}^{(i)}_{\ell m}\bm{g}_{\ell m}
+{\cal F}^{(i)}_{\ell m}\bm{f}_{\ell m}
\right]\,,
\label{eq:Tharm}
\end{eqnarray}
where $\bm{a}_{0\,\ell m}$, $\bm{a}_{\ell m}\,,\cdots$ are 
tensor harmonics defined by 
(3.2-11) in \cite{Nakano:2007cj}. 

The tensor harmonics can be classified 
into even and odd parities from the above expressions. 
Even parity modes are defined by the parity $(-1)^\ell$ 
under the transformation $(\theta,\phi)\to(\pi-\theta,\phi+\pi)$, 
while odd parity modes are by the parity $(-1)^{\ell+1}$. 
Using the orthogonality of the above tensor harmonics, 
we can derive the coefficient of the corresponding tensor harmonics expansion. 
For example, 
\begin{eqnarray}
{\cal A}^{(1)}_{0\,\ell m}(t,r) = \int \bm{T}^{(1)} 
\cdot \bm{a}_{0\,\ell m}^{*}\, d\Omega 
= \int \delta^{\mu \alpha} \delta^{\nu \beta} 
T^{(1)}_{\mu\nu} \,a_{0\,\ell m \,\alpha\beta}^{*}\, d\Omega \,,
\end{eqnarray}
where $*$ denotes the complex conjugate, 
$d\Omega=\sin \theta d\theta d\phi$ 
and $\delta^{\mu \alpha}$ 
has the component, $
\delta^{\mu \alpha}=diag(1,\,1,\,1/r^2,\,1/(r^2 \sin^2 \theta))$.

We summarize the Regge-Wheeler-Zerilli formalism 
in the time domain. 
The linearized Hilbert-Einstein equation
is given in \eref{eq:formal1st}. 
Next, we specify that the stress-energy tensor 
in the right hand side of the above equation 
be the one of a point particle moving along a geodesic; 
\begin{eqnarray}
T^{(1)}{}^{\mu\nu}&= \mu \int^{+\infty}_{-\infty} \delta^{(4)}(x-z(\tau))
{dz^{\mu} \over d\tau}{dz^{\nu} \over d\tau}d\tau \nonumber \\
&= \mu \, U^0 {dz^{\mu} \over dt}{dz^{\nu} \over dt}
{\delta(r-R(t)) \over r^2} \delta ^{(2)}(\Omega-\Omega (t)) \,,
\label{eq:pp-form}
\end{eqnarray}
where we have used the following notations, 
\begin{eqnarray}
z^{\mu}&= z^{\mu}(\tau)
= \{T(\tau),R(\tau),\Theta(\tau),\Phi(\tau)\} \,,
\quad 
U^0 = {dT(\tau) \over d\tau} \,,
\end{eqnarray}
for the particle's orbit. 
This stress-energy tensor is expressed 
in terms of the tensor harmonics as given in \tref{table:SET} 
for the even parity modes. 

\begin{table}
 \caption{The stress-energy tensor for the even parity modes.}
 \label{table:SET}
 \begin{indented}\item[]
  \begin{tabular}{@{}l}
\br
$\displaystyle
{\cal A}^{(1)}_{\ell m}(t,r)=\mu U^0 \left({dR \over dt}\right)^2(r-2M)^{-2}
\delta(r-R(t))Y_{\ell m}^*(\Omega (t))$
\\
$\displaystyle
{\cal A}^{(1)}_{0\,\ell m}(t,r)=\mu U^0 \left(1-{2M \over r}\right)^2r^{-2}
\delta(r-R(t))Y_{\ell m}^*(\Omega (t))$
\\ 
$\displaystyle
{\cal A}^{(1)}_{1\,\ell m}(t,r)=\sqrt{2}i\mu U^0 {dR \over dt}r^{-2}
\delta(r-R(t))Y_{\ell m}^*(\Omega (t))$
\\
$\displaystyle
{\cal B}^{(1)}_{0\,\ell m}(t,r)=[\ell(\ell+1)/2]^{-1/2}i\mu U^0
\left(1-{2M \over r}\right)r^{-1}\delta(r-R(t))dY_{\ell m}^*(\Omega (t))/dt$
\\
$\displaystyle
{\cal B}^{(1)}_{\ell m}(t,r)=[\ell(\ell+1)/2]^{-1/2}\mu U^0
(r-2M)^{-1}{dR \over dt}\delta(r-R(t))dY_{\ell m}^*(\Omega (t))/dt$
\\
%$\displaystyle
%{\cal Q}^{(1)}_{0\,\ell m}(t,r)=-[\ell(\ell+1)/2]^{-1/2}\mu U^0
%\left(1-{2M \over r}\right)r^{-1} \delta(r-R(t))$
%\\
%\qquad \quad $\displaystyle
%\times \left[{1\over \sin \Theta}{\partial Y_{\ell m}^* \over \partial \Phi}
%{d \Theta \over dt}-\sin \Theta{\partial Y_{\ell m}^* \over \partial \Theta}
%{d\Phi \over dt}\right]$ 
%\\
%$\displaystyle
%{\cal Q}^{(1)}_{\ell m}(t,r)=-[\ell(\ell+1)/2]^{-1/2}i\mu U^0 {dR \over dt}
%(r-2M)^{-1}\delta(r-R(t))$
%\\
%\qquad \quad $\displaystyle
%\times \left[{1\over \sin\Theta}{\partial Y_{\ell m}^* \over \partial \Phi}
%{d \Theta \over dt}-\sin \Theta{\partial Y_{\ell m}^* \over \partial \Theta}
%{d\Phi \over dt}\right]$ 
%\\ 
%$\displaystyle
%{\cal D}^{(1)}_{\ell m}(t,r)=-[\ell(\ell+1)(\ell-1)(\ell+2)/2]^{-1/2}i\mu U^0
%\delta(r-R(t))$
%\\
%\qquad \quad $\displaystyle
%\times \left({1 \over 2}\left[({d \Theta \over dt})^2
%-\sin^2\Theta ({d\Phi \over dt})^2\right]
%{1 \over \sin \Theta}X_{\ell m}^*[\Omega(t)]
%-\sin \Theta {d\Phi \over dt}{d \Theta \over dt}
%  W_{\ell m}^*[\Omega(t)]\right)$ 
%\\ 
$\displaystyle
{\cal F}^{(1)}_{\ell m}(t,r)=[\ell(\ell+1)(\ell-1)(\ell+2)/2]^{-1/2}\mu U^0
\delta(r-R(t))$
\\
$\displaystyle
\qquad \quad 
\times \left({d\Phi \over dt}{d \Theta \over dt}X_{\ell m}^*[\Omega(t)]
+{1 \over 2}\left[({d \Theta \over dt})^2
-\sin^2\Theta ({d\Phi \over dt})^2\right]W_{\ell m}^*[\Omega(t)]\right)$ 
\\ 
$\displaystyle
{\cal G}^{(1)}_{\ell m}(t,r)={\mu U^0 \over \sqrt{2}}\delta(r-R(t))
\left[({d \Theta \over dt})^2
+\sin^2\Theta ({d\Phi \over dt})^2\right]Y_{\ell m}^*(\Omega (t))$
\\
\br
  \end{tabular}
 \end{indented}
\end{table}

Substituting \eref{eq:hharm} and \eref{eq:Tharm} into
\eref{eq:formal1st}, 
we obtain the linearized Hilbert-Einstein equation for each harmonic mode. 
Here, we use the RW gauge condition, 
$h_{2\,\ell m}^{(1)}=0$ for the odd part and 
$h_{0\,\ell m}^{(e)(1)}=h_{1\,\ell m}^{(e)(1)}=G_{\ell m}^{(1)}=0$ for the even part. 
From these ten linearized equations, we derive the Regge-Wheeler-Zerilli equations 
and construct the metric perturbations 
from the Regge-Wheeler-Zerilli functions in the RW gauge. 
In the following, we focus on the even parity modes. 

We introduce the following function, 
\begin{eqnarray}
\fl
\psi^{\rm even}_{\ell m}(t,r)=
\frac{2\,r}{\ell(\ell+1)} \Bigl[
K^{(1)}_{\ell m} (t,r)
\nonumber \\ 
+2\,{\frac { ( r-2\,M )  }
{ ( r{\ell}^{2}+r\ell-2\,r+6\,M ) }}
\left(H^{(1)}_{2\,\ell m} ( t,r )
-r\,
{\frac {\partial }{\partial r}}K^{(1)}_{\ell m}(t,r)\right)
\Bigr]
\,.
\label{eq:defpsi-form}
\end{eqnarray}
This function is related to Zerilli's even parity function 
$\psi_{\ell m}^{\rm Z,even}$ as 
\begin{eqnarray}
\fl
\psi_{\ell m}^{\rm Z,even}(t,r)
=
\frac{2}{r{\ell}^{2}+r\ell-2\,r+6\,M}
\left({r}^2{\frac {\partial }{\partial t}}K^{(1)}_{\ell m} ( t,r )
-( r-2\,M ) H^{(1)}_{1\,\ell m} ( t,r ) \right)
\nonumber \\ 
\fl \qquad \qquad \quad 
= 
\frac{\partial}{\partial t} \, \psi^{\rm even}_{\ell m}(t,r)
- \,{\frac {16\,\sqrt {2}\,\pi i\,{r}^{2} (r-2\,M )}
{\ell (\ell + 1)( r{\ell}^{2}+r\ell-2\,r+6\,M )}}\,
{\cal A}^{(1)}_{1\,\ell m}(t,r) \,.
\label{eq:ZerilliDef}
\end{eqnarray}
The function $\psi^{\rm even}_{\ell m}$ obeys the Zerilli equation, 
\begin{eqnarray}
\left[-\frac{\partial^2}{{\partial t}^2} 
+ \frac{\partial^2}{{\partial r^*}^2}-V_\ell^{\rm even}(r) \right]
\psi^{\rm even}_{\ell m}(t,r)=S^{\rm even}_{\ell m}(t,r) \,,
\label{eq:zerilli-form}
\end{eqnarray}
where $r^*=r+2M \ln(r/2M-1)$ and 
\begin{eqnarray}
\fl
V_{\ell}^{{\rm even}}(r)
= 
\frac {r-2M}{{r}^{4} ( r{\ell}^{2}+r\ell-2\,r+6\,M ) ^{2}}
\left( \ell ( \ell+1 )  ( \ell+2 ) ^{2} ( \ell-1 ) ^{2}{r}^{3}
 \right. \nonumber \\  \left. \qquad 
+6\,M ( \ell+2 ) ^{2} ( \ell-1 ) ^{2}{r}^{2}
+36\,{M}^{2} ( \ell+2 )  ( \ell-1 ) r+72\,{M}^{3} \right) \,,
\label{eq:Veven}
\end{eqnarray}
and the source term is given by 
\begin{eqnarray}
\fl
S_{\ell m}^{\rm even}(t,r) 
=
{\frac {16\,\pi \, ( r-2\,M ) ^{2} ( r{\ell}^{2}+r\ell-4\,r+2\,M ) }
{ \ell ( \ell+1 )( r{\ell}^{2}+r\ell-2\,r+6\,M ) r }}{\cal A}^{(1)}_{\ell m} ( t,r ) 
\nonumber \\  \hspace{-15mm}
-{\frac { 16\, \sqrt {2}\,\pi ( r-2\,M )}
{\sqrt {\ell ( \ell+1 )  ( \ell-1 )  ( \ell+2 ) }}}{\cal F}^{(1)}_{\ell m} ( t,r ) 
+{\frac {32\,\pi \, ( r-2\,M ) ^{2}\sqrt {2}}
{ ( r{\ell}^{2}+r\ell-2\,r+6\,M ) \sqrt {\ell ( \ell+1 ) }}}{\cal B}^{(1)}_{\ell m} ( t,r ) 
\nonumber \\ \hspace{-15mm} 
-{\frac {32\,\pi ( r-2\,M ) ^{3}  }
{ ( r{\ell}^{2}+r\ell-2\,r+6\,M ) \ell ( \ell+1 ) }}\,{\frac {\partial }{\partial r}}
{\cal A}^{(1)}_{\ell m} ( t,r )
-\biggl\{
16\,\pi \,r ( {\ell}^{4}{r}^{2}+2\,{r}^{2}{\ell}^{3}-5\,{r}^{2}{\ell}^{2}
\nonumber \\ \hspace{-15mm} 
+16\,r{\ell}^{2}M-6\,{r}^{2}\ell+16\,r\ell M+8\,{r}^{2}-68\,rM
+108\,{M}^{2} ) /\bigl[ ( \ell+1 ) \ell ( r{\ell}^{2}+r\ell
\nonumber \\ \hspace{-15mm}
-2\,r+6\,M ) ^{2}\bigr]
\biggr\} {\cal A}^{(1)}_{0\,\ell m} ( t,r ) 
+{\frac { 32\,\pi ( r-2\,M ) {r}^{2}}
{ ( r{\ell}^{2}+r\ell-2\,r+6\,M ) \ell ( \ell+1 ) }}
\,{\frac {\partial }{\partial r}}{\cal A}^{(1)}_{0\,\ell m} ( t,r ) 
\nonumber \\ \hspace{-15mm} 
+{\frac {32\,\sqrt {2}\, \pi( r-2\,M ) ^{2} }
{ ( r{\ell}^{2}+r\ell-2\,r+6\,M ) \ell ( \ell+1 ) }}{\cal G}^{(1)}_{\ell m} ( t,r ) 
\,.
\label{eq:Seven}
\end{eqnarray}
Here $T_{\mu\nu}^{(1)}{}^{;\nu}=0$ have been used to simplify the source term. 
Using the function $\psi^{\rm even}_{\ell m}$, 
the four coefficients for the metric perturbations 
in the RW gauge are expressed as 
\begin{eqnarray}
\fl
K^{(1)RW}_{\ell m} ( t,r ) 
= \frac{1}{2}\,\biggl\{
( {\ell}^{4}{r}^{2}+2\,{r}^{2}{\ell}^{3}
-{r}^{2}{\ell}^{2}+6\,r{\ell}^{2}M-2\,{r}^{2}\ell+6\,r\ell M 
\nonumber \\ \quad 
-12\,rM+24\,{M}^{2} )
/ \left[( r{\ell}^{2}+r\ell-2\,r+6\,M ) {r}^{2}\right]
\biggr\}\,\psi_{\ell m}^{\rm (even)}(t,r)
\nonumber \\ 
+{\frac { ( r-2\,M )  }{r}}\,{\frac {\partial }{\partial r}}\psi_{\ell m}^{\rm (even)}(t,r)
-{\frac {32\,\pi {r}^{3}}{\ell ( \ell+1 )  
( r{\ell}^{2}+r\ell-2\,r+6\,M ) }}\,{\cal A}^{(1)}_{0\,\ell m} ( t,r ) \,, 
\nonumber \\
\fl
H^{(1)RW}_{2\,\ell m} ( t,r ) 
= -\frac{1}{2}\,{\frac { ( r{\ell}^{2}+r\ell-2\,r+6\,M ) }{r-2\,M}}\, 
K^{(1)RW}_{\ell m} ( t,r )
+r{\frac {\partial }{\partial r}}K^{(1)RW}_{\ell m} ( t,r )
\nonumber \\ 
+\frac{1}{4}\,{\frac {\ell ( \ell+1 ) ( r{\ell}^{2}+r\ell-2\,r+6\,M )  }
{ ( r-2\,M ) r}} \,\psi_{\ell m}^{\rm (even)}(t,r) \,, 
\nonumber \\
\fl
H^{(1)RW}_{0\,\ell m} ( t,r ) 
=H^{(1)RW}_{2\,\ell m} ( t,r ) +16\,{\frac {\pi {r}^{2}\sqrt {2}}
{\sqrt {\ell ( \ell+1 )  ( \ell-1 )  ( \ell+2 ) }}}\,{\cal F}^{(1)}_{\ell m} ( t,r )  \,, 
\nonumber \\
\fl
H^{(1)RW}_{1\,\ell m} ( t,r ) 
= 2\,{\frac {r ( r-3\,M )  }{ ( r-2\,M ) \ell ( \ell+1 ) }}
\,{\frac {\partial }{\partial t}}K^{(1)RW}_{\ell m} ( t,r )
+2\,{\frac {{r}^{2}}
{\ell ( \ell+1 ) }}{\frac {\partial ^{2}}{\partial t\partial r}}K^{(1)RW}_{\ell m} ( t,r ) 
\nonumber \\ 
-2\,{\frac {r}{\ell ( \ell+1 ) }}
\,{\frac {\partial }{\partial t}}H^{(1)RW}_{2\,\ell m} ( t,r ) 
+{\frac {8\,i\pi  \sqrt {2}{r}^{2}}{\ell ( \ell+1 ) }}\,{\cal A}^{(1)}_{1\,\ell m} ( t,r ) \,.
\label{eq:recE}
\end{eqnarray}

%%%%%%%%%%%%%%%%%%%%%%%%%%%%%%%%%%%%%%%%%%%%%%%%%%%%%%%%%%%%%%%%%%%%%%
\section{Second order source terms}\label{app:sost} 
%%%%%%%%%%%%%%%%%%%%%%%%%%%%%%%%%%%%%%%%%%%%%%%%%%%%%%%%%%%%%%%%%%%%%%

In this appendix, we show the 
regularized source for the second order Zerilli equation in 
\eref{eq:regZeq} from the contribution of the 
$(\ell=2)\cdot(\ell=2)$ coupling explicitly. 
We can also derive the source from the $(\ell=0)\cdot(\ell=2)$ coupling 
in the same way, but the expression is so long that we do not have the space to show it here. 
The regularized second order source \eref{eq:S22Zreg}
from the $(\ell=2)\cdot(\ell=2)$ coupling has 
the following six factors 
\begin{eqnarray}
\fl
{}^{G(2,2)}S^{out}_{20}(t,r) 
= {\displaystyle \frac {2}{7}} \sqrt{5}\,(r - 2\,M) 
\left[ {\vrule height1.38em width0em depth1.38em}
 \right. \!  \!  - {\displaystyle \frac {3\,(8\,r^{2} + 12\,M\,r + 7\,M^{2})\,(r - 2\,M)
}{r^{2}\,(2\,r + 3\,M)}}\,
\Psi_{20}^{out}(t,r) 
\nonumber \\ \hspace{-18mm} \quad
\times  \left({\frac {\partial ^{2}}{\partial t\,\partial r}}\,\Psi_{20}^{out}(t,r)\right)
+ {\displaystyle \frac {3\,(2\,r^{2} - M^{2})\,(r - 2\,M)}{r\,(2\,r + 3\,M)}}
 \,\left({\frac {\partial ^{2}}{\partial r^{2}}}
\,\Psi_{20}^{out}(t,r)\right)
\nonumber \\ \hspace{-18mm} \quad 
\times 
\left({\frac {\partial }{\partial t}}\,\Psi_{20}^{out}(t,r)\right) 
- {\displaystyle \frac {(18\,r^{3} - 4\,r^{2}\,M - 33\,r\,M^{2} - 48\,M^{3})}
{r^{2}\,(2\,r + 3\,M)}} 
\,\left({\frac {\partial }{\partial t}}\,\Psi_{20}^{out}(t,r)\right)
\nonumber \\ \hspace{-18mm} \quad 
\times \left({\frac {\partial }{\partial r}}\,\Psi_{20}^{out}(t,r)\right) 
- (r - 2\,M)^{2}\,
\left({\frac {\partial ^{3}}{\partial t\,\partial r^{2}}}\,\Psi_{20}^{out}(t,r)\right)
\,\left({\frac {\partial }{\partial r}}\,\Psi_{20}^{out}(t,r)\right) 
- 3\,(112\,r^{5}
\nonumber \\ \hspace{-18mm} \quad 
 + 480\,r^{4}\,M
 + 692\,r^{3}\,M^{2} 
+ 762\,r^{2}\,M^{3} + 441\,r\,M^{4} + 144\,M^{5}) 
\,\left({\frac {\partial }{\partial t}}\,\Psi_{20}^{out}(t,r) \right)
\nonumber \\ \hspace{-18mm}\quad 
\times \Psi_{20}^{out}(t,r)\left/ 
{\vrule height0.41em width0em depth0.41em} \right. \!  \!  
(r^{3}\,(2\,r + 3\,M)^{3}) 
- {\displaystyle \frac {(r - 2\,M)\,(3\,r - 7\,M)}{r}} 
\,\left({\frac {\partial }{\partial r}}\,\Psi_{20}^{out}(t,r)\right)
\nonumber \\ \hspace{-18mm}\quad
\times \left({\frac {\partial ^{2}}{\partial t\,\partial r}}\,\Psi_{20}^{out}(t,r)\right) 
+ {\displaystyle \frac {3\,(r - 2\,M)^{2}\,M}{r\,(2\,r + 3\,M)}} \,
\left({\frac {\partial ^{3}}{\partial t\,\partial r^{2}}}\,\Psi_{20}^{out}(t,r)\right)
\,\Psi_{20}^{out}(t,r) 
\nonumber \\ \hspace{-18mm}
+ (r - 2\,M)^{2}\,
\left({\frac {\partial ^{3}}{\partial r^{3}}}\,\Psi_{20}^{out}(t,r)\right)\,
\left({\frac {\partial }{\partial t}}\,\Psi_{20}^{out}(t,r)\right) 
\! \! \left. {\vrule height1.38em width0em depth1.38em} \right] 
\left/ 
{\vrule height0.41em width0em depth0.41em} \right. \!  \! 
[\sqrt{\pi }\,(2\,r + 3\,M)\,r^{2}] 
\,, \\
\fl
{}^{G(2,2)}S^{in}_{20}(t,r) = {\displaystyle \frac {2}{7}} \sqrt{5}\,(r - 2\,M) 
\left[ {\vrule height1.38em width0em depth1.38em} 
\right. \!  \! {\displaystyle \frac {3\,(r - 2\,M)^{2}\,M}{r\,(2\,r + 3\,M)}}\,
\Psi_{20}^{in}(t,r)\,\left({\frac {\partial ^{3}}{\partial t\,\partial r^{2}}}\,
\Psi_{20}^{in}(t,r)\right)  
\nonumber \\ \hspace{-18mm}
- (r - 2\,M)^{2}\,\left({\frac {\partial }{\partial r}}\,
\Psi_{20}^{in}(t,r)\right)\,
\left({\frac {\partial ^{3}}{\partial t\,\partial r^{2}}}
\,\Psi_{20}^{in}(t,r)\right) - (18\,r^{3} - 4\,r^{2}\,M - 33\,r\,M^{2} 
\nonumber \\ \hspace{-18mm} \quad 
- 48\,M^{3}) /(r^{2}\,(2\,r + 3\,M))
\,\left({\frac {\partial }{\partial t}}\,\Psi_{20}^{in}(t,r)\right)
\,\left({\frac {\partial }{\partial r}}\,\Psi_{20}^{in}(t,r)\right)  
\nonumber \\ \hspace{-18mm}
- {\displaystyle \frac {(r - 2\,M)\,(3\,r - 7\,M)}{r}} \,
\left({\frac {\partial ^{2}}{\partial t\,\partial r}}\,\Psi_{20}^{in}(t,r)\right)
\,\left({\frac {\partial }{\partial r}}\,\Psi_{20}^{in}(t,r)\right) 
- 3\,(112\,r^{5} + 480\,r^{4}\,M 
\nonumber \\ \hspace{-18mm} \quad 
+ 692\,r^{3}\,M^{2} 
+ 762\,r^{2}\,M^{3} + 441\,r\,M^{4} + 144\,M^{5}) /
(r^{3}\,(2\,r + 3\,M)^{3}) \Psi_{20}^{in}(t,r) 
\nonumber \\ \hspace{-18mm}\quad \times
\left({\frac {\partial }{\partial t}}\,\Psi_{20}^{in}(t,r)\right)\,
- \frac{3\,(8\,r^{2} + 12\,M\,r + 7\,M^{2})
\,(r - 2\,M) }{r^{2}\,(2\,r + 3\,M)} \,
\left({\frac {\partial ^{2}}{\partial t\,\partial r}}\,
\Psi_{20}^{in}(t,r)\right)
\nonumber \\ \hspace{-18mm} \quad \times 
\Psi_{20}^{in}(t,r) + {\displaystyle \frac {3\,(2\,r^{2} - M^{2})\,(r - 2\,M)
}{r\,(2\,r + 3\,M)}}\,
\left({\frac {\partial ^{2}}{\partial r^{2}}}\,\Psi_{20}^{in}(t,r)\right)\,
\left({\frac {\partial }{\partial t}}\,\Psi_{20}^{in}(t,r)\right)  
\nonumber \\ \hspace{-18mm}
+ (r - 2\,M)^{2}\,\left({\frac {\partial ^{3}}{\partial r^{3}}}\,
\Psi_{20}^{in}(t,r)\right)
\,\left({\frac {\partial }{\partial t}}\,\Psi_{20}^{in}(t,r)\right) 
\! \! \left. {\vrule height1.38em width0em depth1.38em} \right] 
 \left/ 
{\vrule height0.41em width0em depth0.41em} \right. \!  \! 
[\sqrt{\pi } (2\,r + 3\,M)\,r^{2}] 
\,, \\
\fl
{}^{G(2,2)}S^{\delta}_{20}(t,r) 
= {\displaystyle \frac {4}{7}} \sqrt{5}\,\sqrt{\pi }\,
\mu \,Y_{20}^*(0,0) \Biggl[ 
- \frac{dR(t)}{dt}\,( - R(t) + 2\,M)^{2}\,(2\,M + 3\,E^{2}\,R(t) - R(t))
\nonumber \\ \hspace{-18mm}\quad \times 
\,{\left({\frac {\partial ^{3}}{\partial r^{3}}}\,\Psi_{20}^{in}(t,r)\right)
\vrule \lower 7pt \hbox{$\displaystyle \, r=R(t)$}} 
- 2 \frac{dR(t)}{dt}( - 38\,R(t)\,M^{3} 
- 48\,M^{4} - 40\,R(t)^{2}\,E^{2}\,M^{2} 
\nonumber \\ \hspace{-18mm} \quad 
+ 282\,M^{3}\,R(t)\,E^{2} 
+ 288\,M^{4}\,E^{2} 
+ 171\,R(t)^{2}\,M^{2} + 66\,M\,E^{2}\,R(t)^{3} 
+ 12\,R(t)^{4} 
\nonumber \\ \hspace{-18mm}\quad 
- 94\,R(t)^{3}\,M - 64\,R(t)^{4}\,E^{2}) 
\times {\left({\frac {\partial }{\partial r}}\,\Psi_{20}^{in}(t,r)\right)
\vrule \lower 7pt \hbox{$\displaystyle \, r=R(t)$}} 
\left/ {\vrule height0.44em width0em depth0.44em} \right. \!  \! 
(R(t)\,(2\,R(t) + 3\,M)^{2}) 
\nonumber \\ \hspace{-18mm}
+ 6\,\frac{dR(t)}{dt}( - 48\,E^{2}\,R(t)^{5} 
+ 8\,R(t)^{5} + 288\,M^{5}\,E^{2} + 546\,R(t)\,M^{4}\,E^{2} 
+ 52\,R(t)^{4}\,M\,E^{2} 
\nonumber \\ \hspace{-18mm}\quad 
- 60\,R(t)^{4}\,M - 41\,R(t)^{2}\,M^{3} 
+ 288\,R(t)^{2}\,M^{3}\,E^{2} 
+ 104\,R(t)^{3}\,M^{2} 
+ 36\,M^{5} 
\nonumber \\ \hspace{-18mm}\quad 
- 8\,R(t)^{3}\,M^{2}\,E^{2}) \Psi_{20}^{in}(t, \,R(t)) 
\left/ {\vrule height0.44em width0em depth0.44em} \right. \!  \! 
(R(t)^{2}\,(2\,R(t) + 3\,M)^{3}) - ( - R(t) + 2\,M)
\nonumber \\ \hspace{-18mm}\quad
\times 
(18\,M^{3} + 5\,R(t)\,M^{2} 
+ 6\,M^{2}\,R(t)\,E^{2} + 14\,M\,R(t)^{2}\,E^{2} - 43\,R(t)^{2}\,M + 18\,R(t)^{3}
\nonumber \\ \hspace{-18mm} \quad
- 12\,R(t)^{3}\,E^{2} ) 
{\left({\frac {\partial ^{2}}{\partial r\,\partial t}}\,
\Psi_{20}^{in}(t,r)\right)\vrule \lower 7pt \hbox{$\displaystyle \, r=R(t)$}}
\left/ {\vrule height0.44em width0em depth0.44em} \right. \!  \! 
(R(t)\,(2\,R(t) + 3\,M))+ (216\,M^{5} 
\nonumber \\ \hspace{-18mm} \quad 
+ 162\,R(t)\,M^{4} 
- 72\,R(t)\,M^{4}\,E^{2} - 150\,R(t)^{2}\,M^{3}\,E^{2} + 165\,R(t)^{2}\,M^{3} 
- 4\,R(t)^{5} 
\nonumber \\ \hspace{-18mm}\quad 
- 386\,R(t)^{3}\,M^{2} + 36\,R(t)^{3}\,M^{2}\,E^{2} 
- 108\,R(t)^{4}\,M\,E^{2} + 126\,R(t)^{4}\,M 
\nonumber \\ \hspace{-18mm}\quad 
+ 56\,E^{2}\,R(t)^{5}) 
{\left({\frac {\partial }{\partial t}}\,\Psi_{20}^{in}(t,r)\right)
\vrule \lower 7pt \hbox{$\displaystyle \, r=R(t)$}} 
\left/ {\vrule height0.44em width0em depth0.44em} \right. \!  \! 
(R(t)^{2}\,(2\,R(t) + 3\,M)^{2}) 
\nonumber \\ \hspace{-18mm}
- {\displaystyle \frac {1}{2}} \,( - R(t) + 2\,M)^{2}\,
(10\,M + 6\,E^{2}\,R(t) - 5\,R(t))
{\left({\frac {\partial ^{3}}{\partial r^{2}\,\partial t}}\,\Psi_{20}^{in}(t,r)\right)
\vrule \lower 7pt \hbox{$\displaystyle \, r=R(t)$}}  
\nonumber \\ \hspace{-18mm}
+ ( - R(t) + 2\,M)\,\frac{dR(t)}{dt}(48\,M^{3} + 26\,R(t)\,M^{2} 
+ 66\,M^{2}\,R(t)\,E^{2} + 34\,M\,R(t)^{2}\,E^{2} 
\nonumber \\ \hspace{-18mm}\quad 
- 13\,R(t)^{2}\,M + 12\,R(t)^{3}\,E^{2} 
- 6\,R(t)^{3})
{\left({\frac {\partial ^{2}}{\partial r^{2}}}\,\Psi_{20}^{in}(t,r)\right)
\vrule \lower 7pt \hbox{$\displaystyle \, r=R(t)$}}
\nonumber \\ \hspace{-18mm}\quad 
\left/ {\vrule height0.44em width0em depth0.44em} \right. \!  \! 
(R(t)\,(2\,R(t) + 3\,M)) 
- \frac{dR(t)}{dt}\,( - R(t) + 2\,M)^{2}\,
(2\,M + 3\,E^{2}\,R(t) - R(t))
\nonumber \\ \hspace{-18mm}\quad \times 
{\left({\frac {\partial ^{3}}{\partial r^{3}}}\,\Psi_{20}^{out}(t,r)\right)
\vrule \lower 7pt \hbox{$\displaystyle \, r=R(t)$}} 
- {\displaystyle \frac {1}{2}} \,( - R(t) + 2\,M)^{2}\,
(10\,M + 6\,E^{2}\,R(t) - 5\,R(t))
\nonumber \\ \hspace{-18mm}\quad \times 
{\left({\frac {\partial ^{3}}{\partial r^{2}\,\partial t}}\,\Psi_{20}^{out}(t,r)\right)
\vrule \lower 7pt \hbox{$\displaystyle \, r=R(t)$}} 
- 4\,\pi \frac{dR(t)}{dt}\,\mu \,Y_{20}^*(0,0)(810\,M^{6} + 1113\,M^{5}\,R(t) 
\nonumber \\ \hspace{-18mm}\quad 
+ 972\,M^{5}\,R(t)\,E^{2} 
+ 1710\,M^{4}\,R(t)^{2}\,E^{2} + 1161\,M^{4}\,R(t)^{2} 
+ 2124\,M^{3}\,R(t)^{3}\,E^{2} 
\nonumber \\ \hspace{-18mm}\quad 
- 2352\,M^{3}\,R(t)^{3} 
+ 328\,M^{2}\,R(t)^{4} - 1296\,M^{2}\,R(t)^{4}\,E^{2} - 96\,M\,R(t)^{5}\,E^{2} 
\nonumber \\ \hspace{-18mm}\quad 
+ 120\,M\,R(t)^{5} + 32\,R(t)^{6} - 288\,E^{2}\,R(t)^{6})E 
\nonumber \\ \hspace{-18mm}\quad 
\left/ {\vrule height0.44em width0em depth0.44em} \right. \!  \! 
(R(t)\,(2\,R(t) + 3\,M)^{4}\,( - R(t) + 2\,M))\Biggr] 
\left/ {\vrule height0.44em width0em depth0.44em} \right. \!  \! 
[E R(t)^{2}\,(2\,R(t) + 3\,M)] 
\,, \\
\fl
{}^{G(2,2)}S^{\delta'}_{20}(t,r) 
= {\displaystyle \frac {4}{7}} \mu \,\sqrt{5}\,\sqrt{\pi }
\,Y_{20}^*(0,0) \Biggl[ - ( - R(t) + 2\,M) 
( - 10\,M^{2} + 6\,E^{2}\,M^{2}
 + 17\,R(t)\,M 
\nonumber \\ \hspace{-18mm}\quad 
- 4\,M\,E^{2}\,R(t) 
- 6\,R(t)^{2} + 4\,E^{2}\,R(t)^{2})\,\frac{dR(t)}{dt} 
{\left({\frac {\partial }{\partial r}}\,\Psi_{20}^{in}(t,r)\right)
\vrule \lower 7pt \hbox{$\displaystyle \, r=R(t)$}}
\nonumber \\ \hspace{-18mm} \quad 
\left/ {\vrule height0.44em width0em depth0.44em} \right. \!  \!  
(2\,R(t) + 3\,M)
+ 6\,( - R(t) + 2\,M)
(3\,E^{2}\,M^{3} + 2\,M^{3} 
+ 6\,M^{2}\,R(t)\,E^{2} - 3\,R(t)\,M^{2} 
\nonumber \\ \hspace{-18mm}\quad 
+ 4\,M\,R(t)^{2}\,E^{2} 
+ 5\,R(t)^{2}\,M 
+ 4\,R(t)^{3}\,E^{2} - 2\,R(t)^{3}) 
\frac{dR(t)}{dt}\Psi_{20}^{in}(t, \,R(t)) 
\nonumber \\ \hspace{-18mm}\quad 
\left/ {\vrule height0.44em width0em depth0.44em} \right. \!  \! 
(R(t)\,(2\,R(t) + 3\,M)^{2}) 
+ ( - R(t) + 2\,M)^{2}\,(2\,E^{2}\,R(t) 
- 3\,R(t) + 6\,M)
\nonumber \\ \hspace{-18mm}\quad \times 
{\left({\frac {\partial ^{2}}
{\partial r\,\partial t}}\,\Psi_{20}^{in}(t,r)\right)
\vrule \lower 7pt \hbox{$\displaystyle \, r=R(t)$}} 
+ 2\,\frac{dR(t)}{dt}\,( - R(t) + 2\,M)^{2}\,
(E^{2}\,R(t) - R(t) + 2\,M)
\nonumber \\ \hspace{-18mm}\quad \times 
{\left({\frac {\partial ^{2}}{\partial r^{2}}}\,\Psi_{20}^{in}(t,r)\right)
\vrule \lower 7pt \hbox{$\displaystyle \, r=R(t)$}} 
+  ( - R(t) + 2\,M)\,(3\,M^{2} + 6\,R(t)\,M - 2\,R(t)^{2})
\nonumber \\ \hspace{-18mm} \quad \times 
\,(2\,E^{2}\,R(t) - 3\,R(t) 
+ 6\,M) {\left({\frac {\partial }{\partial t}}\,\Psi_{20}^{in}(t,r)\right)
\vrule \lower 7pt \hbox{$\displaystyle \, r=R(t)$}}
\left/ {\vrule height0.44em width0em depth0.44em} \right. \!  \! 
(R(t)\,(2\,R(t) + 3\,M)) \Biggr]
\nonumber \\ \hspace{-18mm}
\left/ {\vrule height0.44em width0em depth0.44em} \right. \!  \! 
[E\,R(t)^{2}\,(2\,R(t) + 3\,M)] 
\,, \\
\fl
{}^{T(2,2)}S^{\delta}_{20}(t,r) 
= {\displaystyle \frac {8}{7}} \sqrt{5}\,\sqrt{\pi }\,\mu 
\,(R(t) - 2\,M)\,Y_{20}^*(0,0) 
\left[ {\vrule height1.23em width0em depth1.23em} \right. \!  \!  
- 2(24\,M^{3} + 10\,R(t)\,M^{2} 
\nonumber \\ \hspace{-18mm} \quad
+ 6\,M^{2}\,R(t)\,E^{2} + 9\,M\,R(t)^{2}\,E^{2} - 7\,R(t)^{2}\,M 
+ R(t)^{3}\,E^{2} 
- 2\,R(t)^{3})\frac{dR(t)}{dt}
\nonumber \\ \hspace{-18mm}\quad \times 
{\left({\frac {\partial }{\partial r}}\,\Psi_{20}^{in}(t,r)\right)
\vrule \lower 7pt \hbox{$\displaystyle \, r=R(t)$}}
\left/ {\vrule height0.44em width0em depth0.44em} \right. \!  \! 
(R(t)\,(2\,R(t) + 3\,M)) + 6\,\frac{dR(t)}{dt}(24\,M^{4} 
\nonumber \\ \hspace{-18mm} \quad 
+ 6\,M^{3}\,R(t)\,E^{2} 
+ 36\,R(t)\,M^{3} + 21\,R(t)^{2}\,M^{2} + 13\,R(t)^{2}\,E^{2}\,M^{2}
+ 10\,R(t)^{3}\,M 
\nonumber \\ \hspace{-18mm}\quad 
+ 8\,M\,E^{2}\,R(t)^{3} + 6\,R(t)^{4}\,E^{2} 
- 4\,R(t)^{4}) 
\Psi_{20}^{in}(t, \,R(t)) 
\left/ {\vrule height0.44em width0em depth0.44em} \right. \!  \! 
(R(t)^{2}\,(2\,R(t) + 3\,M)^{2}) 
\nonumber \\ \hspace{-18mm}
+ (E^{2}\,R(t) - R(t) + 2\,M)\,
( - R(t) + 2\,M)\,{\left({\frac {\partial ^{2}}
{\partial r\,\partial t}}\,\Psi_{20}^{in}(t,r)\right)
\vrule \lower 7pt \hbox{$\displaystyle \, r=R(t)$}} 
\nonumber \\ \hspace{-18mm}
+ (E^{2}\,R(t) - R(t) + 2\,M)\,\frac{dR(t)}{dt}
\,( - R(t) + 2\,M)
{\left({\frac {\partial ^{2}}{\partial r^{2}}}\,\Psi_{20}^{in}(t,r)\right)
\vrule \lower 7pt \hbox{$\displaystyle \, r=R(t)$}} 
\nonumber \\ \hspace{-18mm}
- {\displaystyle \frac {6\,(E^{2}\,R(t) 
- R(t) + 2\,M)\,(R(t)^{2} + R(t)\,M + M^{2})}
{R(t)\,(2\,R(t) + 3\,M)}} 
{\left({\frac {\partial }{\partial t}}\,\Psi_{20}^{in}(t,r)\right)
\vrule \lower 7pt \hbox{$\displaystyle \, r=R(t)$}}
\! \! \left. {\vrule height1.23em width0em depth1.23em} \right] 
 \nonumber \\ \hspace{-18mm}
\left/ {\vrule height0.44em width0em depth0.44em} \right. \!  \! 
[ER(t)^{3}\,(2\,R(t) + 3\,M)] 
\,, \\
\fl
{}^{T(2,2)}S^{\delta'}_{20}(t,r) 
= {\displaystyle \frac {8}{7}} 
\sqrt{5}\,\sqrt{\pi }\,\mu 
\,(R(t) - 2\,M)^{2}
\,Y_{20}^*(0,0) 
\left[ {\vrule height1.23em width0em depth1.23em} \right. \!  \!  
\frac{dR(t)}{dt}\,( - R(t) + 2\,M)\,
\nonumber \\ \hspace{-18mm} \quad 
\times {\left({\frac {\partial }{\partial r}}\,\Psi_{20}^{in}(t,r)\right)
\vrule \lower 7pt \hbox{$\displaystyle \, r=R(t)$}} 
- \frac{dR(t)}{dt}{\displaystyle \frac {6\,
\,(R(t)^{2} + R(t)\,M + M^{2})}
{R(t)\,(2\,R(t) + 3\,M)}} \,\Psi_{20}^{in}(t, \,R(t)) 
\! \! \left. {\vrule height1.23em width0em depth1.23em} \right] 
\nonumber \\ \hspace{-18mm}
\left/ {\vrule height0.44em width0em depth0.44em} \right. \!  \! 
(E\,R(t)^{3}\,(2\,R(t) + 3\,M)) \,.
\end{eqnarray}
Here, we have used \eref{eq:localQ} 
for $\Psi^{\Theta}_{20}$ and 
its derivatives at the particle trajectory. 
And for the homogeneous solutions, 
$\Psi_{20}^{out}$ and $\Psi_{20}^{in}$ have been used 
to write the above source terms.

%%%%%%%%%%%%%%%%%%%%%%%%%%%%%%%%%%%%%%%%%%%%%%%%%%%%%%%%%%%%%%%%%%%%%%
\section{Second order gauge transformation} \label{app:SOGT}
%%%%%%%%%%%%%%%%%%%%%%%%%%%%%%%%%%%%%%%%%%%%%%%%%%%%%%%%%%%%%%%%%%%%%%

In this appendix, we deal with first and second order gauge transformations. 
In order two obtain the second order waveform, 
it is necessary to derive the second order metric perturbations
in an asymptotic flat (AF) gauge. 
The Regge-Wheeler-Zerilli formalism 
that we have employed in the RW gauge 
is not asymptotically flat. 
Therefore, we will focus on the gauge transformation from the RW gauge 
to an AF gauge. We also need 
to discuss the first order gauge transformation to an AF gauge 
simultaneously. 

Here, we consider the following gauge transformation~\cite{Mukhanov:1996ak,Bruni:1996im}. 
\begin{eqnarray}
x^{\mu}_{RW} &\to& x^{\mu}_{AF} 
= x^{\mu}_{RW} + \xi^{(1)\mu} \left(x^{\alpha}\right)
+\frac{1}{2} \left[\xi^{(2)\mu}\left(x^{\alpha}\right)
+\xi^{(1)\nu} \xi^{(1)\mu}{}_{,\nu}\left(x^{\alpha}\right)\right] \,,
\end{eqnarray}
where comma "," in the index indicates the partial derivative 
with respect to the background coordinates, 
and $\xi^{(1)\mu}$ and $\xi^{(2)\mu}$ are generators of 
the first and second order gauge transformations, respectively. 
Then, the metric perturbations 
changes as 
\begin{eqnarray}
h_{RW \mu \nu}^{(1)} &\to h_{AF \mu \nu}^{(1)}=
h_{RW \mu \nu}^{(1)} - {\cal L}_{\xi^{(1)}} g_{\mu \nu} \,,
\label{eq:genGT1} \\
h_{RW \mu \nu}^{(2)} &\to h_{AF \mu \nu}^{(2)}=
h_{RW \mu \nu}^{(2)} 
-\frac{1}{2} {\cal L}_{\xi^{(2)}} g_{\mu \nu}
+\frac{1}{2} {\cal L}_{\xi^{(1)}}^2 g_{\mu \nu}
-{\cal L}_{\xi^{(1)}} h_{RW \mu \nu}^{(1)} \,.
\label{eq:genGT2}
\end{eqnarray}
Next, we discuss the $(\ell=2)\cdot(\ell=2)$ 
and $(\ell=0)\cdot(\ell=2)$ parts separately.

%%%%%%%%%%%%%%%%%%%%%%%%%%%%%%%%%%%%%%%%%%%%%%%%%%%%%%%%%%%%%%%%%%%%%%
\subsection{First order $\ell=2$ mode and second order $(\ell=2)\cdot(\ell=2)$ part} 
%%%%%%%%%%%%%%%%%%%%%%%%%%%%%%%%%%%%%%%%%%%%%%%%%%%%%%%%%%%%%%%%%%%%%%

In this paper, we have used only the even parity mode, 
therefore a generator of the gauge transformation for $\ell=2$, $m=0$ modes 
can be written as
\begin{eqnarray}
\xi^{(i)\mu}_{\ell=2} &=
\biggl\{V_0^{(i)}(t,r) Y_{20}(\theta,\phi),\,V_1^{(i)}(t,r) Y_{20}(\theta,\phi),\,
\nonumber \\ 
& \qquad V_2^{(i)}(t,r) \partial_{\theta} Y_{20}(\theta,\phi),\,
V_2^{(i)}(t,r) \frac{\partial_{\phi} Y_{20}(\theta,\phi)}{\sin^2 \theta}
\biggl\} \,,
\end{eqnarray}
where $i=1$ and $2$ denote the first and second perturbative order, respectively. 
There are three degrees of gauge freedom 
for each perturbative order. 

The gauge transformation of the metric perturbations is explicitly given as follows: 
For the first order metric perturbations, we find 
\begin{eqnarray}
\fl
H_{0\,20}^{(1)AF}(t,r) = H_{0\,20}^{(1)RW}(t,r)
+ 2\,{\frac {\partial }{\partial t}}{V_0}^{(1)} \left( t,r \right) +2\,{
\frac {M}{r \left( r-2\,M \right) }}{V_1}^{(1)} \left( t,r \right)  \,, 
\nonumber \\ 
\fl
H_{1\,20}^{(1)AF}(t,r) = H_{1\,20}^{(1)RW}(t,r)
+ {\frac { \left( r-2\,M \right) }{r}}{\frac {\partial }{\partial r}}{V_0}^{(1)}
 \left( t,r \right) -{\frac {r }{r-2\,M}}{\frac {\partial }{\partial t}}{
V_1}^{(1)} \left( t,r \right) \,,
\nonumber \\ 
\fl
H_{2\,20}^{(1)AF}(t,r) = H_{2\,20}^{(1)RW}(t,r)
-2\,{\frac {\partial }{\partial r}}{V_1}^{(1)} \left( t,r \right) 
+2\,{\frac {M }{r \left( r-2\,M \right) }}{V_1}^{(1)} \left( t,r \right) \,,
\nonumber \\ 
\fl
K_{20}^{(1)AF}(t,r) = K_{20}^{(1)RW}(t,r)
-{\frac {2}{r}}{V_1}^{(1)} \left( t,r \right)  \,,
\nonumber \\ 
\fl
h_{0\,20}^{(e)(1)AF}(t,r) = 
{\frac { \left( r-2\,M \right)  }{r}}{V_0}^{(1)} \left( t,r \right)
-{r}^{2}{\frac {\partial }{\partial t}}{V_2}^{(1)} \left( t,r \right) \,,
\nonumber \\ 
\fl
h_{1\,20}^{(e)(1)AF}(t,r) = 
-{\frac {r }{r-2\,M}}{V_1}^{(1)} \left( t,r \right)
-{r}^{2}{\frac {\partial }{\partial r}}{V_2}^{(1)} \left( t,r \right) \,,
\nonumber \\ 
\fl
G_{20}^{(1)AF}(t,r) = 
-2\,{V_2}^{(1)} \left( t,r \right) \,.
\end{eqnarray}
For the second order metric perturbations, 
we can calculate the gauge transformation straightforwardly, 
but we obtain very long expressions. 
For example, they can be written formally as 
\begin{eqnarray}
\fl
K_{20}^{(2)AF}(t,r) = K_{20}^{(2)RW}(t,r) 
-\frac {1}{r}{V_1}^{(2)} \left( t,r \right) +  \delta K_{20}^{(2)}(t,r)   \,,
\\
\fl
h_{1\,20}^{(e)(2)AF}(t,r) = 
-{\frac {r }{2(r-2\,M)}}{V_1}^{(2)} \left( t,r \right)
-\frac{{r}^{2}}{2}{\frac {\partial }{\partial r}}{V_2}^{(2)} \left( t,r \right) 
+ \delta h_{1\,20}^{(2)(e)}(t,r) \,.
\\
\fl
G_{20}^{(2)AF}(t,r) = 
-{V_2}^{(1)} \left( t,r \right) + \delta G_{20}^{(2)}(t,r) \,,
\end{eqnarray}
where $\delta K_{20}^{(2)}$, $\delta h_{1\,20}^{(2)(e)}$ 
and $\delta G_{20}^{(2)}$ 
are defined by the tensor harmonics expansion of 
the last two terms in the right hand side 
of \eref{eq:genGT2}. 
This includes only quadratic terms of 
the first order wave-function. 

First, we consider the asymptotic behavior on 
the $\ell=2$ mode of the first order 
metric perturbations in the RW gauge. 
The asymptotic expansion of the wave-function $\psi^{\rm even}_{2 0}$ 
is given by 
\begin{eqnarray}
\fl
\psi^{\rm even}_{2 0}(t,r) = 
\frac{1}{3}\,{\frac {d^{2}}{d{{\it T_r}}^{2}}}F \left( {\it T_r} \right) 
+\left({{\frac {d}{d{\it T_r}}}F \left( {\it T_r} \right) }\right){r^{-1}}
+\left({F \left( {\it T_r} \right) -M{\frac {d}{d{\it T_r}}}F \left( {\it T_r}
 \right) }\right){{r}^{-2}}
\nonumber \\ 
+ \Or(r^{-3})  \,,
\label{eq:asympt1st}
\end{eqnarray}
where we have introduced ${\it T_r} = t-r_*(r)$ for simplicity. 
In the following calculation, 
we need only the leading order contribution with respect to 
the above large $r$ expansion. 
Then, the coefficients of the metric perturbations 
are given, from \eref{eq:recE}, as follows 
\begin{eqnarray}
H_{0\,20}^{(1)RW}(t,r) = H_{2\,20}^{(1)RW}(t,r) = 
\frac{1}{3}\, \left( {\frac {d^{4}}{d{{\it T_r}}^{4}}}F \left( {\it T_r}
 \right)  \right) r + \Or(r^{0}) \,,
\nonumber \\ 
H_{1\,20}^{(1)RW}(t,r) = 
-\frac{1}{3}\, \left( {\frac {d^{4}}{d{{\it T_r}}^{4}}}F \left( {\it T_r}
 \right)  \right) r + \Or(r^{0})\,,
\nonumber \\ 
K_{20}^{(1)RW}(t,r) = 
-\frac{1}{3}\,{\frac {d^{3}}{d{{\it T_r}}^{3}}}F \left( {\it T_r} \right) 
+ \Or(r^{-1})
\,.
\end{eqnarray}

On the other hand, the metric perturbations in an AF gauge 
should behave as 
\begin{eqnarray}
\fl
H_{0\,20}^{(1)AF}(t,r) = H_{1\,20}^{(1)AF}(t,r)=h_{0\,20}^{(e)(1)AF}(t,r)=0 \,.
\quad
H_{2\,20}^{(1)AF}(t,r) = \Or(r^{-3}) \,,
\nonumber \\ 
\fl
h_{1\,20}^{(e)(1)AF}(t,r) = \Or(r^{-1}) \,,
K_{20}^{(1)AF}(t,r) = \Or(r^{-1}) \,,
\quad
G_{20}^{(1)AF}(t,r) = \Or(r^{-1}) \,.
\label{eq:AFgauge}
\end{eqnarray}
This asymptotic behavior will also be considered 
for the second order calculation. 
We find the following gauge transformation to go 
to the AF gauge. 
\begin{eqnarray}
V_0^{(1)}(t,r) = 
-\frac{1}{6}\, \left( {\frac {d^{3}}{d{{\it T_r}}^{3}}}F \left( {\it T_r}
 \right)  \right) r + \Or(r^{0}) \,, 
\nonumber \\
V_1^{(1)}(t,r) = 
-\frac{1}{6}\, \left( {\frac {d^{3}}{d{{\it T_r}}^{3}}}F \left( {\it T_r}
 \right)  \right) r + \Or(r^{0}) \,,
\nonumber \\ 
V_2^{(1)}(t,r) = 
-\frac{1}{6}\,\left(
{{\frac {d^{2}}{d{{\it T_r}}^{2}}}F \left( {\it T_r}
 \right) }\right){r^{-1}} + \Or(r^{-2})
\,.
\end{eqnarray}
The above results are calculated iteratively 
for large $r$ expansion. 
Since the transverse-traceless tensor harmonics 
for the even parity part is $\bm{f}_{\ell m}$ 
in \eref{eq:hharm}, 
the coefficient of the metric perturbations 
related to the gravitational wave is 
$G_{\ell m}^{(1)AF}$. 
This becomes 
\begin{eqnarray}
G_{20}^{(1)AF}(t,r) &= 
\frac{1}{3}\,{\frac {1}{r}}{\frac {d^{2}}{d{{\it T_r}}^{2}}}F \left( {\it T_r}
 \right) + \Or(r^{-2}) \nonumber \\ 
&= 
{\frac {1}{r}}\psi^{\rm even}_{2 0}(t,r) 
+ \Or(r^{-2}) \,.
\end{eqnarray}
with the use of \eref{eq:asympt1st} 

Next, we discuss the second perturbative order. 
When we treat the second order metric perturbations 
from the $(\ell=2)\cdot(\ell=2)$ coupling, in practice, 
we calculate ${\tilde \chi}_{20}^{\rm Z}$ numerically 
instead of $\chi_{20}^{\rm Z}(t,r)$, where ${\tilde \chi}_{20}^{\rm Z}$ 
has been considered in \eref{eq:formalreg} as 
\begin{eqnarray}
{\tilde \chi}_{20}^{\rm Z}(t,r) 
&=& \chi_{20}^{\rm Z}(t,r) - \chi_{20}^{{\rm reg},(2,2)}(t,r)
- \chi_{20}^{{\rm reg},(0,2)}( t,r ) \,,
\end{eqnarray}
where $\chi_{20}^{{\rm reg},(0,2)}$ is the $(\ell=0)\cdot(\ell=2)$ 
contribution, to be discussed in the next subsection. 
Here, to derive the gravitational wave amplitude 
for the second perturbative order, 
we also need to obtain the coefficient $G_{2 0}^{(2)AF}$ 
in an AF gauge as in the first order case. 

The asymptotic expansion of ${\tilde \chi}_{20}^{\rm Z}$ is 
\begin{eqnarray}
{\tilde \chi}_{20}^{\rm Z}(t,r) &=& 
\frac{1}{3}\,{\frac {d^{3}}{d{{\it T_r}}^{3}}}F_2 \left( {\it T_r} \right) 
+ \Or(r^{-1})  
\,.
\end{eqnarray}
Here, we may consider only the leading order contribution with respect to 
large $r$ expansion in the same manner as the first order calculation. 
The ${\tilde \chi}_{20}^{\rm Z}$ contribution 
to the waveform is derived by the same method 
as that for the first perturbative order. 

First, we obtain $\partial K_{20}^{(2)RW}/\partial t$ 
in the RW gauge from \eref{eq:2ndKRW}
as 
\begin{eqnarray}
\fl
{\frac {\partial }{\partial t}}K_{20}^{(2)RW} \left( t,r \right) = 
-\frac{1}{3}\,{\frac {d^{4}}{d{{\it T_r}}^{4}}}F_2 \left( {\it T_r} \right) 
+ \frac{\sqrt {5}}{18\sqrt {\pi }}\,
\left( {\frac {d^{4}}{d{{\it T_r}}^{4}}}F
 \left( {\it T_r} \right)  \right) {\frac {d^{3}}{d{{\it T_r}}^{3}}}F
 \left( {\it T_r} \right) 
\nonumber \\  + \Or(r^{-1}) \,,
\end{eqnarray}
where the second term in the right hand side of the above equation 
arises from the regularization function $\chi_{20}^{{\rm reg},(2,2)}$ and 
the ${\cal A}^{(1)}_{20}$ and ${\cal B}^{(0)}_{20}$-terms 
in \eref{eq:2ndKRW}. 
Integrating the above equation for $K_{20}^{(2)RW}$
\begin{eqnarray}
\fl
K_{20}^{(2)RW} \left( t,r \right) 
= -\frac{1}{3}\,{\frac {d^{3}}{d{{\it T_r}}^{3}}}F_2 \left( {\it T_r} \right) 
+ \frac{\sqrt {5}}{36\sqrt {\pi }}\,\left( {\frac {d^{3}}{d{{\it T_r}}^{3}}}F
 \left( {\it T_r} \right)  \right) ^{2} 
+ \Or(r^{-1}) \,.
\end{eqnarray}
The second order gauge transformation of 
$K_{20}^{(2)}$ is given by 
\begin{eqnarray}
K_{20}^{(2)AF}(t,r) = K_{20}^{(2)RW}(t,r) 
-\frac {1}{r}{V_1}^{(2)} \left( t,r \right) +  \delta K_{20}^{(2)}(t,r)  \,,
\label{eq:2ndGTK}
\end{eqnarray}
where, $\delta K_{20}^{(2)}$ is defined by the tensor harmonics expansion of
$(1/2) {\cal L}_{\xi^{(1)}}^2 g_{\mu \nu}
-{\cal L}_{\xi^{(1)}} h_{RW \mu \nu}^{(1)}$ in \eref{eq:genGT2} and 
derived as 
\begin{eqnarray}
\fl
 \delta K_{20}^{(2)}(t,r)
=  
{\frac {\sqrt {5}}{252\sqrt {\pi }}}\,\left[
\left( {\frac {d^{4}}{d{{\it T_r}}^{4}}}F
 \left( {\it T_r} \right)  \right) {\frac {d^{2}}{d{{\it T_r}}^{2}}}F
 \left( {\it T_r} \right) 
- 2\left( {\frac {d^{3}}{d{{\it T_r}}^{3}}}F \left( {\it T_r} \right)  \right) ^{2}
\right]
\nonumber \\
 + \Or(r^{-1}) \,.
\end{eqnarray}
From the above results and the AF gauge condition for $K^{(2)}_{20}$ 
in \eref{eq:AFgauge}, 
in order to remove the $O(r^0)$ terms, 
the second order gauge transformation ${V_1}^{(2)}$ is
\begin{eqnarray}
\fl
{V_1}^{(2)} \left( t,r \right) = 
-\frac{1}{3}\, \left( {\frac {d^{3}}{d{{\it T_r}}^{3}}}F_2 \left( {\it T_r}
 \right)  \right) r
\nonumber \\ \hspace{-12mm}
+{\frac {\sqrt {5}}{252\sqrt {\pi }}}\,
\left[ 5\,
 \left( {\frac {d^{3}}{d{{\it T_r}}^{3}}}F \left( {\it T_r} \right) 
 \right) ^{2}
+ \left( {\frac {d^{4}}{d{{\it T_r}}^{4}}}F \left( {\it T_r
} \right)  \right) {\frac {d^{2}}{d{{\it T_r}}^{2}}}F \left( {\it T_r}
 \right)  \right] r
 + \Or(r^0) \,.
\label{eq:V12}
\end{eqnarray}
Here, we note that it is sufficient to consider 
the leading order with respect to large $r$ to derive the second order waveform. 

Next, we consider to derive ${V_2}^{(2)}$ from the condition 
of $h_{1\,20}^{(e)(2)AF}$. 
The second order gauge transformation is given by 
\begin{eqnarray}
\fl
h_{1\,20}^{(e)(2)AF}(t,r) = 
-{\frac {r }{2(r-2\,M)}}{V_1}^{(2)} \left( t,r \right)
-\frac{{r}^{2}}{2}{\frac {\partial }{\partial r}}{V_2}^{(2)} \left( t,r \right) 
+ \delta h_{1\,20}^{(e)(2)}(t,r) \,.
\label{eq:2ndGTh1e}
\end{eqnarray}
$\delta h_{1\,20}^{(e)(2)}$ is defined by the tensor harmonics expansion of
$(1/2) {\cal L}_{\xi^{(1)}}^2 g_{\mu \nu}
-{\cal L}_{\xi^{(1)}} h_{RW \mu \nu}^{(1)}$ in Eq.~(\ref{eq:genGT2}). 
By considering the asymptotic expansion, we obtain 
\begin{eqnarray}
\fl
\delta h_{1\,20}^{(e)(2)}(t,r) = 
{\frac {\sqrt {5}}{504\sqrt {\pi }}}\,
\left[\left( {\frac {d^{4}}{d{{
\it T_r}}^{4}}}F \left( {\it T_r} \right)  \right) {\frac {d^{2}}{d{{
\it T_r}}^{2}}}F \left( {\it T_r} \right) 
+\left( {\frac {d^{3}}{d{{\it T_r}}^{3}}}F
 \left( {\it T_r} \right)  \right) ^{2}
\right] r 
\nonumber \\ 
+ \Or(r^0) 
\,.
\end{eqnarray}
Then, ${V_2}^{(2)}$ is calculated from the above value and the result 
for ${V_1}^{(2)}$ in \eref{eq:V12} 
with the AF gauge condition in Eq.(\ref{eq:AFgauge}) as 
\begin{eqnarray}
\fl
{\frac {\partial }{\partial r}}{V_2}^{(2)} \left( t,r \right) = 
\frac{1}{3}\,{\frac {d^{3}}{d{{\it T_r}}^{3}}}F_2 \left( {\it T_r} \right) \,r^{-1}
-{\frac {\sqrt {5} }{63\sqrt {\pi }}}
\,\left( {\frac {d^{3}}{d{{\it T_r}}^{3}}}
F \left( {\it T_r} \right)  \right) ^{2} r^{-1}
+ \Or(r^{-2}) 
\nonumber \\ \fl \qquad \qquad \quad
= - {\frac {\partial }{\partial t}}{V_2}^{(2)} \left( t,r \right) + \Or(r^{-2}) 
 \,.
\label{eq:V22}
\end{eqnarray}
In the last line, we have used the definition of 
the retarded time, $T_r=t-r_*(r)$. 

From the above results, 
we can consider the metric perturbations related to 
the gravitational wave amplitude, i.e., $G_{20}^{(2)AF}$. 
The gauge transformation is given by 
\begin{eqnarray}
G_{20}^{(2)AF}(t,r) = 
-{V_2}^{(1)} \left( t,r \right) + \delta G_{20}^{(2)}(t,r) \,.
\label{eq:2ndgtG}
\end{eqnarray}
Here, $\delta G_{20}^{(2)}$ is defined by the tensor harmonics expansion of
$(1/2) {\cal L}_{\xi^{(1)}}^2 g_{\mu \nu}
-{\cal L}_{\xi^{(1)}} h_{RW \mu \nu}^{(1)}$ in Eq.~(\ref{eq:genGT2}), 
and found to be
\begin{eqnarray}
\fl
 \delta G_{20}^{(2)}(t,r)
&=  {\frac {\sqrt {5}}{126\sqrt {\pi }}}
\,\left( {\frac {d^{2}}{d{{\it T_r}}^{2}}}F
 \left( {\it T_r} \right)  \right) 
\left({\frac {d^{3}}{d{{\it T_r}}^
{3}}}F \left( {\it T_r} \right)\right) \,r^{-1} 
+ \Or(r^{-2}) \,.
\label{eq:deltaG202}
\end{eqnarray}
Inserting \eref{eq:V22} and \eref{eq:deltaG202} into \eref{eq:2ndgtG}, 
we obtain 
\begin{eqnarray}
\fl 
{\frac {\partial }{\partial t}} G_{20}^{(2)}(t,r) = 
\frac{1}{3}\,{\frac {d^{3}}{d{{\it T_r}}^{3}}}F_2 \left( {\it T_r} \right) r^{-1}
\nonumber \\ \hspace{-18mm}
+{\frac {\sqrt {5}}{126\sqrt {\pi }}}\,
\left[ -
\left( {\frac {d^{3}}{d{{\it T_r}}^{3}}}F \left( {\it T_r} \right)  \right) ^{2}
+
\left( {\frac {d^{4}}{d{{\it T_r}}^{4}}}F \left( {\it T_r} \right)  \right)
 {\frac {d^{2}}{d{{\it T_r}}^{2}
}}F \left( {\it T_r} \right) 
\right] r^{-1}
+ \Or(r^{-2}) 
 \,.
\end{eqnarray}
The gravitational waveform is now (by use of 
$\psi^{\rm even}_{2 0}$ and ${\tilde \chi}_{20}^{\rm Z}$) 
\begin{eqnarray}
\fl 
{\frac {\partial }{\partial t}} G_{20}^{(2)}(t,r) = 
\frac{1}{r}{\tilde \chi}_{20}^{\rm Z}(t,r) 
\nonumber \\ \hspace{-18mm}
+\frac{\sqrt {5}}{14\sqrt {\pi }}\frac{1}{r}\,
\left[-
\left( {\frac {\partial}{\partial t
}}\psi^{\rm even}_{2 0}(t,r)  \right) ^{2}
+\psi^{\rm even}_{2 0}(t,r) 
{\frac {\partial^{2}}{\partial t^{2}}}\psi^{\rm even}_{2 0}(t,r) 
\right]
+ \Or(r^{-2}) 
\,.
\label{eq:FINAL}
\end{eqnarray}
If we consider higher order corrections with respect 
to the $1/r$ expansion, we can show that all metric components satisfy 
the asymptotic flat gauge condition in \eref{eq:AFgauge}. 

%%%%%%%%%%%%%%%%%%%%%%%%%%%%%%%%%%%%%%%%%%%%%%%%%%%%%%%%%%%%%%%%%%%%%%
\subsection{Second order $(\ell=0)\cdot(\ell=2)$ part} 
%%%%%%%%%%%%%%%%%%%%%%%%%%%%%%%%%%%%%%%%%%%%%%%%%%%%%%%%%%%%%%%%%%%%%%

We have already discussed the first order perturbations for the $\ell=0$ mode
in \sref{sec:firstL0}. In this paper, we use the 
first order $\ell=0$ metric perturbation given by \eref{eq:L=0inN}. 
This satisfies the AF gauge condition in \eref{eq:AFgauge}. 
Therefore, it is not necessary to consider the first order gauge transformation 
of the $\ell=0$ mode in \eref{eq:genGT1} and \eref{eq:genGT2}, and 
we will focus on the second perturbative order 
related to the $(\ell=0)\cdot(\ell=2)$ coupling. 

We discuss the gravitational wave amplitude 
for the second perturbative order which arises from 
the $(\ell=0)\cdot(\ell=2)$ coupling 
by using the same method as in the case of the $(\ell=2)\cdot(\ell=2)$ part. 
Note that we have already discussed the contribution from 
${\tilde \chi}_{20}^{\rm Z}$ and $\chi_{20}^{\rm reg,(2,2)}$ in 
the second order wave function of \eref{eq:formalreg}. 
In the following, we consider the $\chi_{20}^{\rm reg,(0,2)}$ 
and the contribution of the last term in the right hand side of \eref{eq:genGT2}, 
i.e., $-{\cal L}_{\xi^{(1)}} h_{RW \mu \nu}^{(1)}$ 
where $\xi^{(1)}$ is the generator of the gauge transformations for $\ell=2$, 
and we use \eref{eq:L=0inN} as the $\ell=0$ mode of $h_{RW \mu \nu}^{(1)}$. 

The regularization function $\chi_{20}^{\rm reg,(0,2)}$ is given in 
\eref{eq:Reg20}. 
This function behaves $\Or(r^{-4})$ for large $r$, 
therefore we expect that $\chi_{20}^{\rm reg,(0,2)}$ does not contribute 
to the second order gravitational waveform at infinity. 

In the same way as for the $(\ell=2)\cdot(\ell=2)$ part, 
we consider the gauge transformation of $K_{20}^{(2)}$ in \eref{eq:2ndGTK}. 
Here, we obtain 
\begin{eqnarray}
K_{20}^{(2)RW}(t,r) +  \delta K_{20}^{(2)}(t,r) = O(r^{-3}) \,,
\end{eqnarray}
where $K_{20}^{(2)RW}$ and $\delta K_{20}^{(2)}$ arise 
from $\chi_{20}^{\rm reg,(0,2)}$ and 
the tensor harmonics expansion of 
$-{\cal L}_{\xi^{(1)}} h_{RW \mu \nu}^{(1)}$ in \eref{eq:genGT2}, 
respectively. 
This is already asymptotically flat, 
therefore we do not need any further gauge transformation, ${V_1}^{(2)}$ 
for the $(\ell=0)\cdot(\ell=2)$ part. 
\begin{eqnarray}
{V_1}^{(2)} \left( t,r \right) &=&  \Or(r^{-2}) \,.
\end{eqnarray}
In \eref{eq:2ndGTh1e}, we have 
\begin{eqnarray}
\delta h_{1\,20}^{(2)(e)}(t,r)
=  \Or(r^{-3}) \,,
\end{eqnarray}
where $\delta h_{1\,20}^{(2)(e)}$ is defined by 
the tensor harmonics expansion of 
$-{\cal L}_{\xi^{(1)}} h_{RW \mu \nu}^{(1)}$ in \eref{eq:genGT2}.
From this equation, we also conclude 
\begin{eqnarray}
{\frac {\partial }{\partial r}}{V_2}^{(2)} \left( t,r \right) = 
 \Or(r^{-4}) 
\,.
\end{eqnarray}

Hence, there is no contribution 
from the $(\ell=0)\cdot(\ell=2)$ coupling to the second order gravitational 
wave, except for ${\tilde \chi}_{20}^{\rm Z}$. This means that 
if we obtain ${\tilde \chi}_{20}^{\rm Z}$ in the numerical calculation, 
we can obtain the gravitational wave amplitude 
for the second perturbative order by using \eref{eq:FINAL}.

%================================================

\end{document}